\RequirePackage{fix-cm}
\documentclass[twocolumn,epjc3]{svjour3}  
\pdfoutput=1

\RequirePackage{amsmath} 
\RequirePackage{graphicx}

\DeclareMathAlphabet{\scr}{U}{rsfs}{m}{n}
\RequirePackage[title]{appendix}
\RequirePackage{amsfonts}
\RequirePackage{amssymb}
\RequirePackage{booktabs}
\RequirePackage{pdfpages}
\RequirePackage{float}
\RequirePackage[utf8]{inputenc}

\usepackage[caption=false]{subfig} 

\usepackage{url}
\usepackage{units}

\usepackage{cite}
\usepackage[colorlinks,citecolor=blue,urlcolor=blue,linkcolor=blue,breaklinks=breakall]{hyperref}
\usepackage{cleveref} 

\crefname{appsec}{Appendix}{Appendices}

\newcommand{\x}{\ensuremath{\mathbf{x}}}
\DeclareMathOperator*{\argmax}{arg\,max}
\newcommand{\ie}{\textit{i.e.}}
\newcommand{\eg}{\textit{e.g.}}

\journalname{Eur. Phys. J. C}

\begin{document}

\title{Signal mixture estimation for degenerate heavy Higgses using a deep
neural network }

\author{Anders Kvellestad\thanksref{addr1,addr2}
	\and Steffen Maeland\thanksref{e2, addr3}
	\and Inga Str{\"u}mke\thanksref{e3, addr3}
}

\institute{Department of Physics, University of Oslo, N-0316 Oslo, Norway \label{addr1}
    \and Department of Physics, Imperial College London, Blackett Laboratory, 
         Prince Consort Road, London SW7 2AZ, UK \label{addr2} 
	\and Department of Physics and Technology, University of Bergen, 
	     N-5020 Bergen, Norway \label{addr3}
}

\thankstext{e2}{e-mail: steffen.maeland@uib.no}
\thankstext{e3}{e-mail: inga.strumke@uib.no}

\date{Received: date / Accepted: date}

\maketitle

\begin{abstract}
	If a new signal is established in future LHC data, a next question will
	be to determine the signal composition, in particular whether the
	signal is due to multiple near-degenerate states.  We investigate the
	performance of a deep learning approach to signal mixture estimation
	for the challenging scenario of a ditau signal coming from a pair of
	degenerate Higgs bosons of opposite $CP$ charge.  This constitutes a
	parameter estimation problem for a mixture model with highly
	overlapping features.  We use an unbinned maximum likelihood fit to a
	neural network output, and compare the results to mixture estimation
	via a fit to a single kinematic variable.  For our benchmark scenarios
	we find a $\sim20\%$ improvement in the estimate uncertainty.
\end{abstract}
%
\section{Introduction}
Machine learning techniques have already proven useful in particle physics,
especially for separating signal from background events in analyses of LHC
data. More recently, \textit{deep learning} methods, such as multi-layer neural
networks, have been shown to perform very well, due to their ability to learn
complex non-linear correlations in high-dimensional
data~\cite{Baldi:2014kfa,Baldi:2014pta,Farbin:2016jll}. In this paper we study
the performance of a deep neural network classifier, but rather than
classifying signal vs.\ background we focus on estimating the mixture of
different signal classes in a dataset. This is motivated by the not-unlikely
scenario where a new (and possibly broad) resonance is discovered in future LHC
data, but limited statistics makes the interpretation difficult, in particular
the question of whether the signal is due to multiple degenerate states. In
such a scenario it will clearly be important to squeeze as much information as
possible from the available data.

While the approach studied here is general, we take a Two-Higgs-Doublet Model
(THDM) as our example scenario. In these models the Higgs sector of the
Standard Model (SM) is extended with an additional $SU(2)$ doublet, predicting
the existence of a pair of charged scalars ($H^\pm$) and three neutral scalars
($h$, $H$, $A$), one of which should be the observed \unit[125]{GeV} Higgs.
Several more extensive frameworks for New Physics predict a Higgs sector with
the structure of a THDM, the prime example being the Minimal Supersymmetric
Standard Model (MSSM). A further motivation for THDMs comes from the fact that
the extended scalar sector can allow for additional sources of $CP$ violation
and a strongly first-order electroweak phase transition, as required for
electroweak
baryogenesis~\cite{Kuzmin:1985mm,Anderson:1991zb,Turok:1990zg,Turok:1991uc}.
For a recent study of this, see~\cite{Andersen:2017ika}.

We associate the light scalar $h$ with the observed \unit[125]{GeV} Higgs and
take the heavier scalars $H$, $A$ and $H^\pm$ to be mass degenerate.  The focus
of our study is on the ditau LHC signal from decays of the neutral states $H$
and $A$, which in this case are indistinguishable save for their opposite $CP$
charges.  Searches for heavy neutral Higgses in ditau final states are carried
out by both the ATLAS and CMS collaborations, see
\cite{Aaboud:2017sjh,Sirunyan:2018zut} for recent results.

The remainder of this paper is structured as follows.
In~\cref{sec:theory_and_motivation} we motivate why it is reasonable to expect
a certain level of mass-degeneracy among the new scalars in THDMs and present
our example THDM scenario. The technical setup for our analysis is given
in~\cref{sec:analysis_setup}. Here we define our signal models, describe the
procedure for Monte Carlo event generation and detail the neural network layout
and training.  In~\cref{sec:phistar_method} we demonstrate $H$/$A$ signal
mixture estimation using the method of fitting a single kinematic variable. The
result serves as our baseline for judging the performance of the deep learning
approach. Our main results are presented in~\cref{sec:network_method}. Here we
estimate the signal mixture via a maximum likelihood fit to the output distribution
from a network trained to separate $H$ and $A$ ditau events. The results are compared
to those from~\cref{sec:phistar_method}. We state our conclusions
in~\cref{sec:conclusion}.
%
\section{Theory and motivation}
\label{sec:theory_and_motivation}
%
The starting point for our study is a THDM scenario where $m_H \approx m_A$.
Our main motivation for this choice is to obtain a challenging test case for
signal mixture estimation. However, there are also physical reasons to expect
the $H$ and $A$ states to have similar masses.  After requiring that the scalar
potential has a minimum in accordance with electroweak symmetry breaking, we
are left with a model with only two mass scales, $v \approx \unit[246]{GeV}$
and a free mass parameter $\mu$, to control the four masses $m_h$, $m_H$, $m_A$
and $m_{H^\pm}$.  From the point of view of the general THDM parameter space,
the least fine-tuned way to align the light state $h$ with SM predictions, as
favoured by LHC Higgs data, is to move towards simultaneous decoupling of the
three heavier states by increasing $\mu$, leaving $v$ to set the scale for $m_h
= \unit[125]{GeV}$~\cite{Bernon:2015qea}. 
This points to a scenario where $|m_H - m_A|\lesssim\unit[100]{GeV}$, and quite
possibly much smaller, depending on the quartic couplings of the scalar
potential.\footnote{A large $H$--$A$ mass difference in this decoupling
scenario relies on $\mathcal{O}(1)$ quartic couplings. We note that when loop
corrections are taken into account, the viability of such scenarios can be
significantly more restricted than what tree-level results
suggest~\cite{Haarr:2016qzq}.}


Further motivation for a small $H$--$A$ mass difference can be found in less
general realisations of THDMs. For the type-II THDM in the MSSM the quartic
couplings are fixed by the squares of the SM gauge couplings, resulting in the
tree-level prediction that $m_H - m_A \lesssim \unit[10]{GeV}$ for \mbox{$m_A
\sim \unit[400]{GeV}$} and $\tan\beta \sim 1$, and decreasing further with
increasing $\tan\beta$ or $m_A$~\cite{Martin:1997ns}.  Another well-motivated
scenario predicting closely degenerate $H$ and $A$ states is the $SO(5)$-based
Maximally Symmetric THDM~\cite{Dev:2014yca}.

When mass degenerate, the $H$ and $A$ appear identical except for their $CP$
charge. If the properties of the light $h$ deviates from SM predictions, this
difference in $CP$ charge can manifest as non-zero $ZZ$ and $WW$ couplings for
$H$, while for the $CP$-odd $A$ the $Zh$ coupling is available. However, these
couplings all vanish in the perfect SM-alignment limit we assume here.  Yet the
$CP$ nature of $H$ and $A$ is still expressed as spin correlations in fermionic
decay modes, impacting the kinematics of subsequent decays. Here we study the
channels $H \rightarrow \tau\tau$ and $A \rightarrow \tau\tau$.  Methods for
reconstructing spin correlations in ditau decays of the \unit[125]{GeV} Higgs
have been investigated in
detail~\cite{dellaquila:1989,Kramer:1993jn,Bower:2002zx,Berge:2015nua},
providing a good baseline for comparison.  The use of neural networks to
optimize $CP$ measurements for the \unit[125]{GeV} state is studied
in~\cite{Jozefowicz:2016kvz,Barberio:2017ngd}. 
%
\subsection{Benchmark scenario}
%
Two-Higgs-Doublet Models are classified in different types based on the
structure of the Yukawa sector.  We choose a benchmark scenario within the
$CP$-conserving lepton-specific THDM, with $m_H = m_A = m_{H^\pm} =
\unit[450]{GeV}$.  In this model, the quarks couple to one of the Higgs
doublets and the leptons to the other. This enables large branching ratios for
$H/A \rightarrow \tau\tau$, even for masses above the \unit[350]{GeV} threshold
for $H/A \rightarrow t\bar{t}$.  

By varying the remaining THDM parameters we can obtain a wide range of ditau
signal strengths for the $H$ and $A$ states at \unit[450]{GeV}.
In~\cref{sec:supplementary_figures} we illustrate how $\sigma(pp \to H) \times
\mathcal{B}(H \to \tau\tau)$ and $\sigma(pp \to A) \times \mathcal{B}(A \to
\tau\tau)$ vary across the high-mass region of the lepton-specific THDM
parameter space.  For $m_H = m_A \approx \unit[450]{GeV}$, we find that the
ditau signal strengths can reach up to $\sigma(pp \to H) \times \mathcal{B}(H
\to \tau\tau) \approx \unit[34]{fb}$ and $\sigma(pp \to A) \times \mathcal{B}
(A \to \tau\tau) \approx \unit[54]{fb}$ in \unit[13]{TeV} proton-proton
collisions.  This includes production via gluon-gluon fusion and bottom-quark
annihilation, with cross sections evaluated at NLO using
\textsf{SusHi~1.6.1}~\cite{Harlander:2012pb,Harlander:2016hcx,Harlander:2002wh,Harlander:2003ai,Actis:2008ug,Harlander:2005rq,Chetyrkin:2000yt}
and branching ratios obtained from \textsf{2HDMC~1.7.0}~\cite{Eriksson:2009ws}.

For comparison, in~\cref{sec:supplementary_figures} we also show the result of
a similar scan of the type-I THDM\@. In this model all fermions couple to only
one of the two Higgs doublets. Compared to the lepton-specific THDM, the ditau
signal in type-I THDM suffers a much stronger suppression from the $H/A
\rightarrow t\bar{t}$ channel. 

As further described in~\cref{sec:analysis_setup,sec:phistar_method}, the
mixture estimation techniques we study require each tau to decay through the
$\tau^\pm \rightarrow \pi^\pm\pi^0\nu$ channel, which has a branching ratio of
25\%.  However, the neural network method we employ can be extended to include
other tau decay modes as well, by implementing the ``impact parameter method''
in~\cite{Berge:2015nua} in addition to the ``$\rho$ decay-plane method'' used
here. 

If we only assume the $\tau^\pm \rightarrow \pi^\pm\pi^0\nu$ decay channel and
an acceptance times efficiency of $5$\%--$10$\% for the signal selection, our
example scenarios predict no more than $\sim100$ signal events for the
anticipated \unit[300]{fb$^{-1}$} dataset at the end of Run 3. However, as the
model scan in~\cref{sec:supplementary_figures} shows, considering slightly
lower benchmark masses can provide an order of magnitude increase in the
predicted cross-section.  Also, extending the method to include more tau decay
channels can greatly increase the statistics available to the analysis
discussed here. 
Still, the large backgrounds in the ditau channel, \eg\ from ``fake QCD taus'',
implies that a signal mixture estimation study for the THDM benchmark scenario 
we present here likely will require the improved statistics of the 
full High-Luminosity LHC dataset. 

We do not include a third mixture component representing ditau backgrounds for 
our benchmark study. Clearly, the inclusion of backgrounds will increase the 
uncertainty in the estimated $H/A \to \tau\tau$ signal mixture. However, as we 
discuss in more detail in~\cref{subsec:results}, the mixture estimate
obtained from the neural network approach we study here is likely to be less 
affected by backgrounds than traditional mixture estimation from fitting a 
single kinematic variable.

For our further discussions we define the parameter $\alpha$ as the ratio of
the $A \rightarrow \tau\tau$ signal strength to the total ditau signal
strength,
\begin{equation}
	\alpha \equiv \frac{\sigma(pp \! \to \! A) \! \times \!
	\mathcal{B}(A \! \to \! \tau\tau)} {\sigma(pp \! \to \!A)
	\! \times \! \mathcal{B}(A \! \to \! \tau\tau) +
	\sigma(pp \! \to \! H) \! \times \! \mathcal{B}
	(H \! \to \! \tau\tau)}.
\label{eq:def_alpha}
\end{equation}
This is the parameter we seek to determine in our signal mixture
estimation.\footnote{In linking this theory quantity directly with the $H$/$A$
event mixture in the datasets we simulate, we make the approximation that the
acceptance times efficiency is equal for $H\rightarrow\tau\tau$ and
$A\rightarrow\tau\tau$ events.} The parameter region of our benchmark scenario
predicts values of $\alpha$ between $0.5$ and $0.7$. To allow for some further
variation in the assumptions, we will in our tests use $\alpha$ values of 0.5,
0.7 and 0.9. 
%
\section{Analysis setup}
\label{sec:analysis_setup}
%
\subsection{Event generation}
\label{subsec:evtgen}
We generate \unit[13]{TeV} $pp$ Monte Carlo events for this study using
\textsf{Pythia~8.219}~\cite{Sjostrand:2006za,Sjostrand:2007gs}. Only
gluon-gluon fusion and bottom-quark annihilation are considered, as these are
the dominant $H$/$A$ production modes at the LHC.\footnote{The magnitudes of
the up-type and down-type Yukawa couplings have the same $\tan\beta$ dependence
in both the lepton-specific and the type-I THDM\@. Gluon-gluon fusion through a
top loop is therefore by far the most important production channel for the
scenarios considered here.} For our analysis we select opposite-sign taus
decaying to $\pi^\pm\pi^0\nu$, which is the decay mode with the highest
branching ratio.  In order to roughly match recent LHC searches for
$H/A\rightarrow\tau\tau$, taus are required to have visible transverse momentum
$p_\mathrm{T}$ larger than \unit[40]{GeV} and pseudorapidity less than 2.1.
Further, we require the taus to be separated by\\ $\Delta R =
\sqrt{{(\Delta\phi)}^2+{(\Delta\eta)}^2} > 0.5$, and that there are no more
than two taus in the event which pass the $p_\mathrm{T}$ selection. Events with
muons or electrons with $p_\mathrm{T} > \unit[20]{GeV}$ are rejected.

Detector effects are taken into account by randomly smearing the directions and
energies of the outgoing pions, following the procedure described
in~\cite{Berge:2015nua}: Each track is deflected by a random polar angle
$\theta$, which is drawn from a Gaussian distribution with width
$\sigma_{\theta}$, so that the smeared track lies within a cone around the true
track direction. For charged pions a value of $\sigma_{\theta} =
\unit[1]{mrad}$ is used, while the energy resolution is $\Delta E/E =
\unit[5]{\%}$. For neutral pions, we use $\sigma_{\theta} =
\unit[0.025/\sqrt{12}]{rad}$ and $\Delta E/E = \unit[10]{\%}$.  To gauge the
impact of such detector effects on our results, we repeat the main analyses
in~\cref{sec:phistar_method,sec:network_method} for simulated data with and
without detector smearing.

%
\subsection{Network input features}
\label{sec:features}
%
\begin{figure*}
	\centering
	\subfloat[][]{%
		\includegraphics[width=0.3\textwidth]{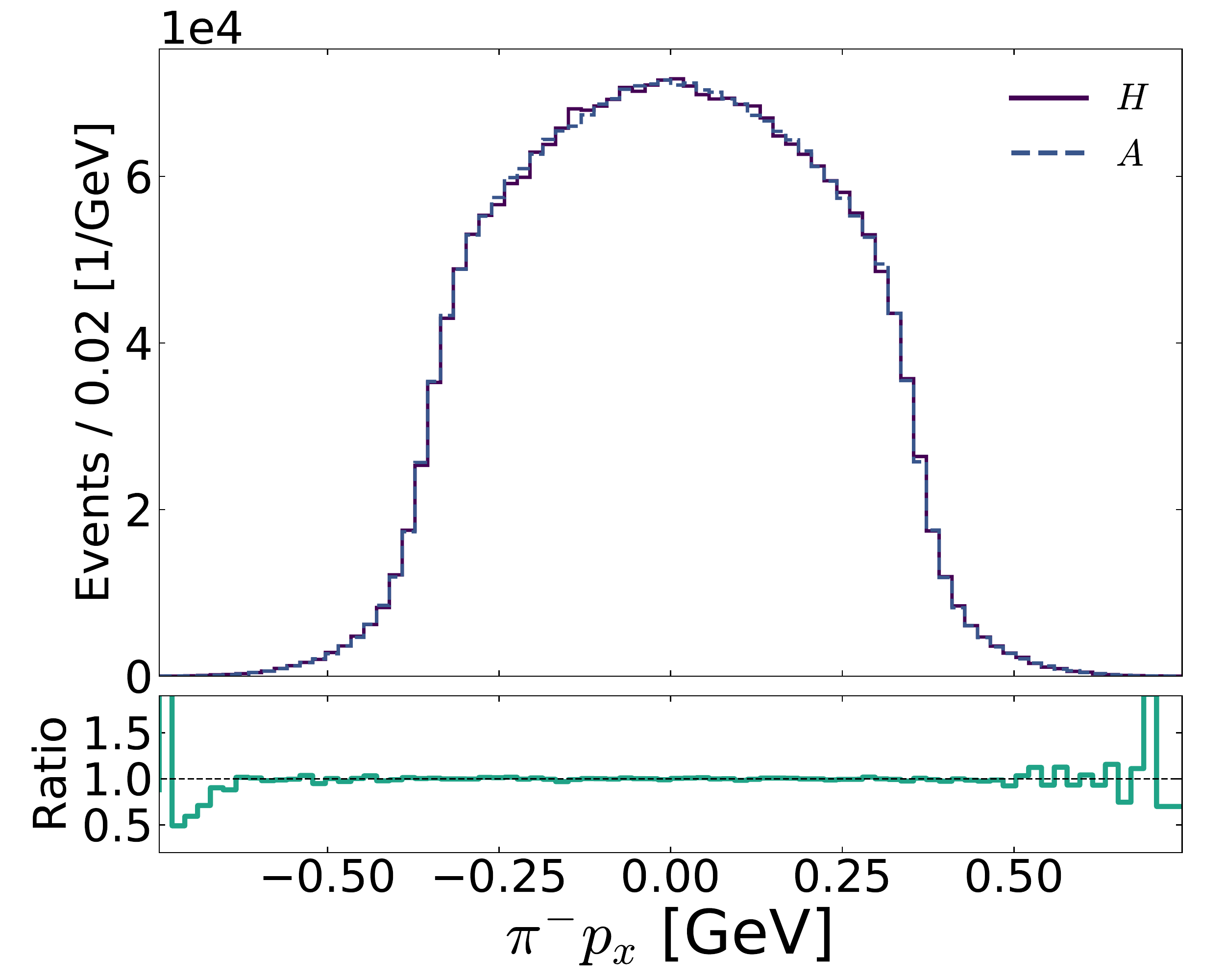}
		\label{fig:piminus_px}}
	\subfloat[][]{%
		\includegraphics[width=0.3\textwidth]{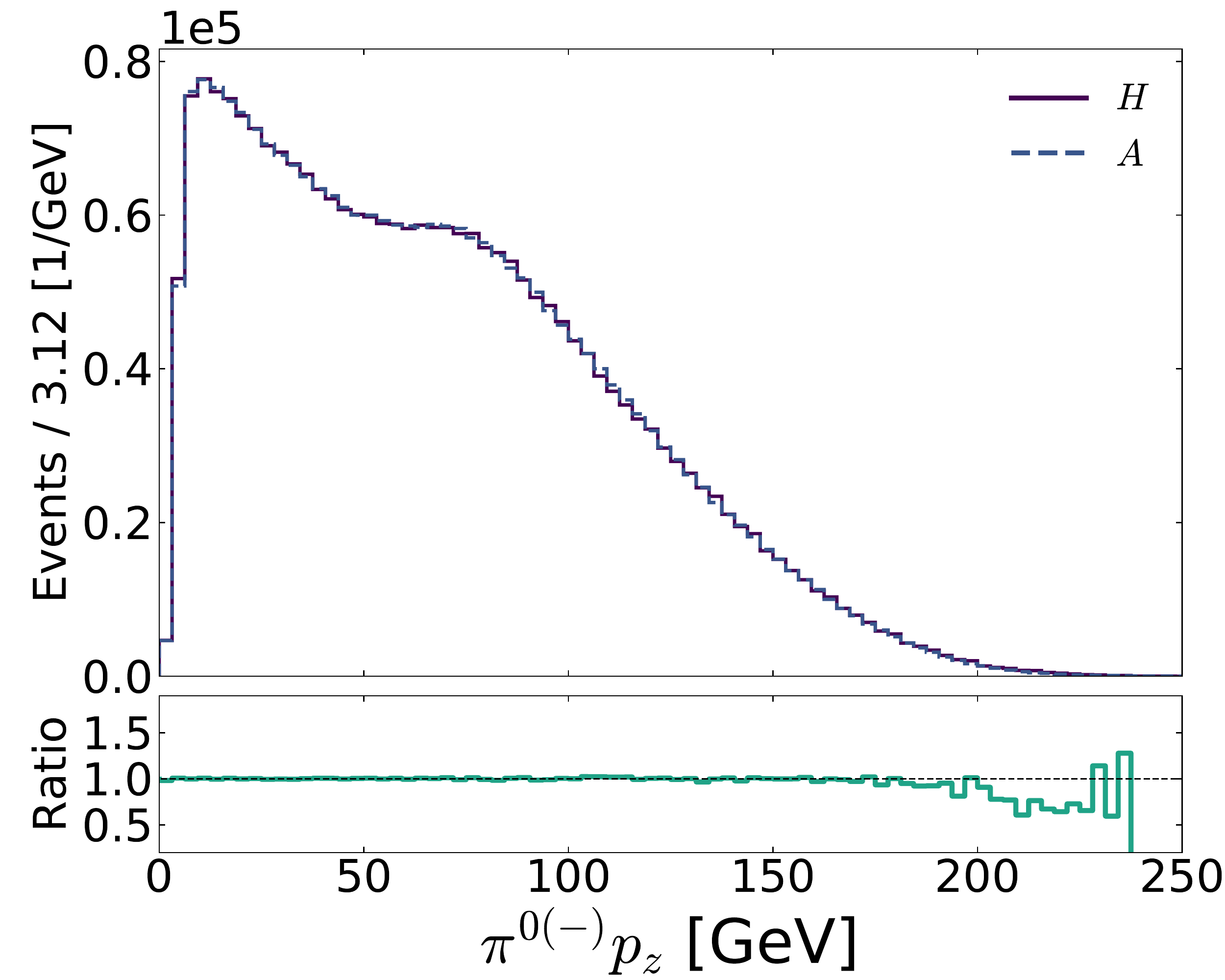}
		\label{fig:pi0minus_pz}}
	\subfloat[][]{%
		\includegraphics[width=0.3\textwidth]{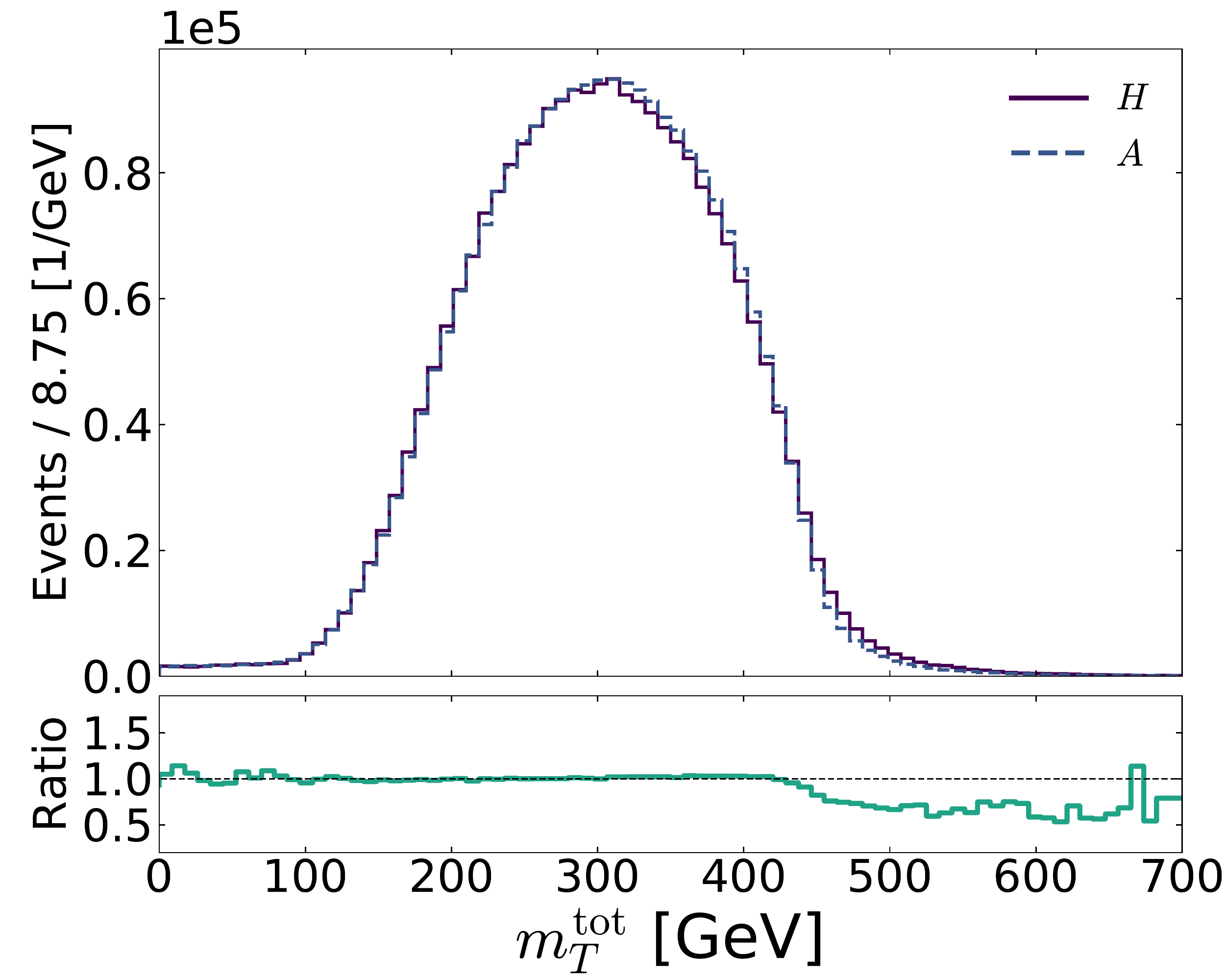}
		\label{fig:transv_mass}}
	\\
	\subfloat[][]{%
		\includegraphics[width=0.3\textwidth]{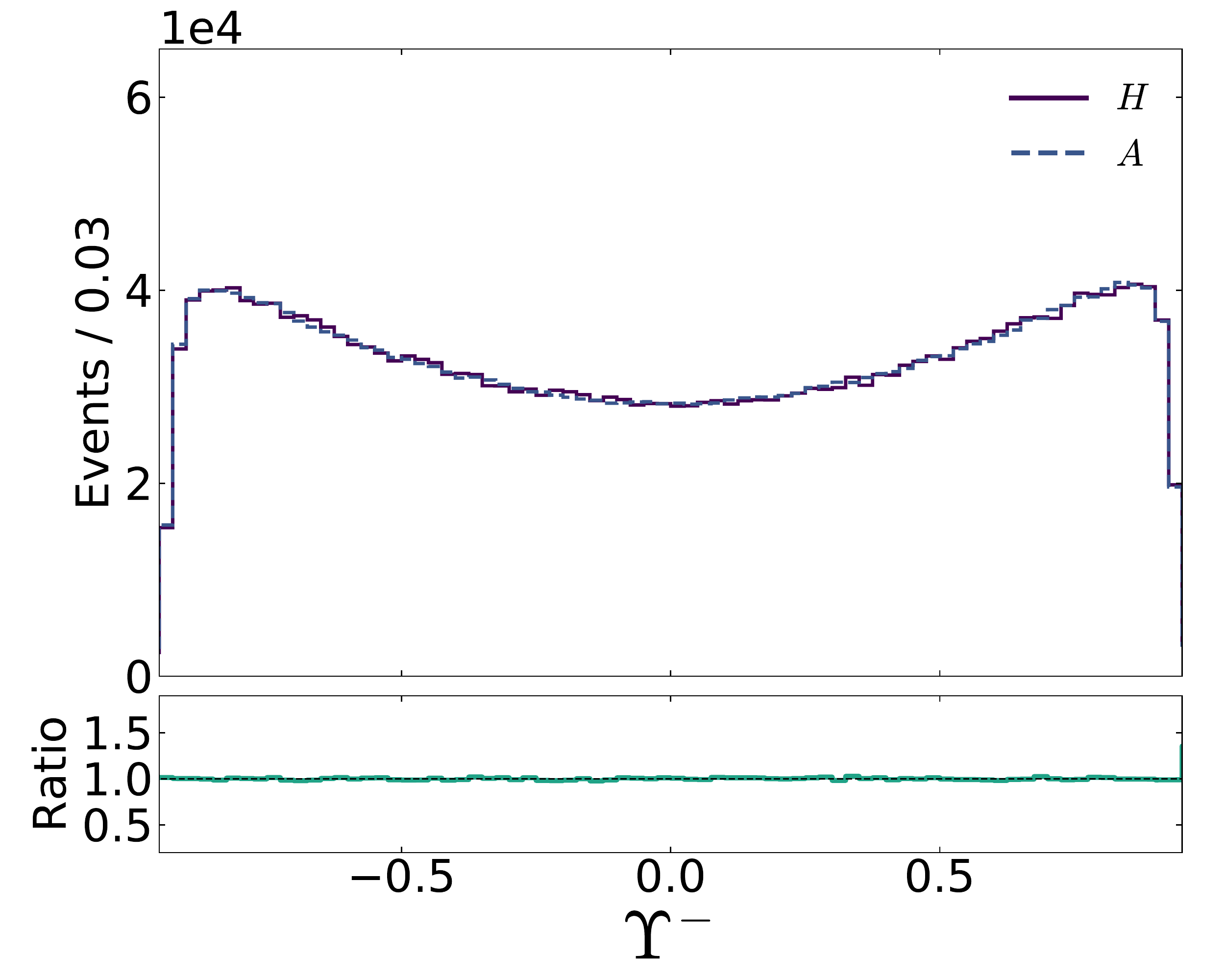}
		\label{fig:upsilon_minus}}
	\subfloat[][]{%
		\includegraphics[width=0.3\textwidth]{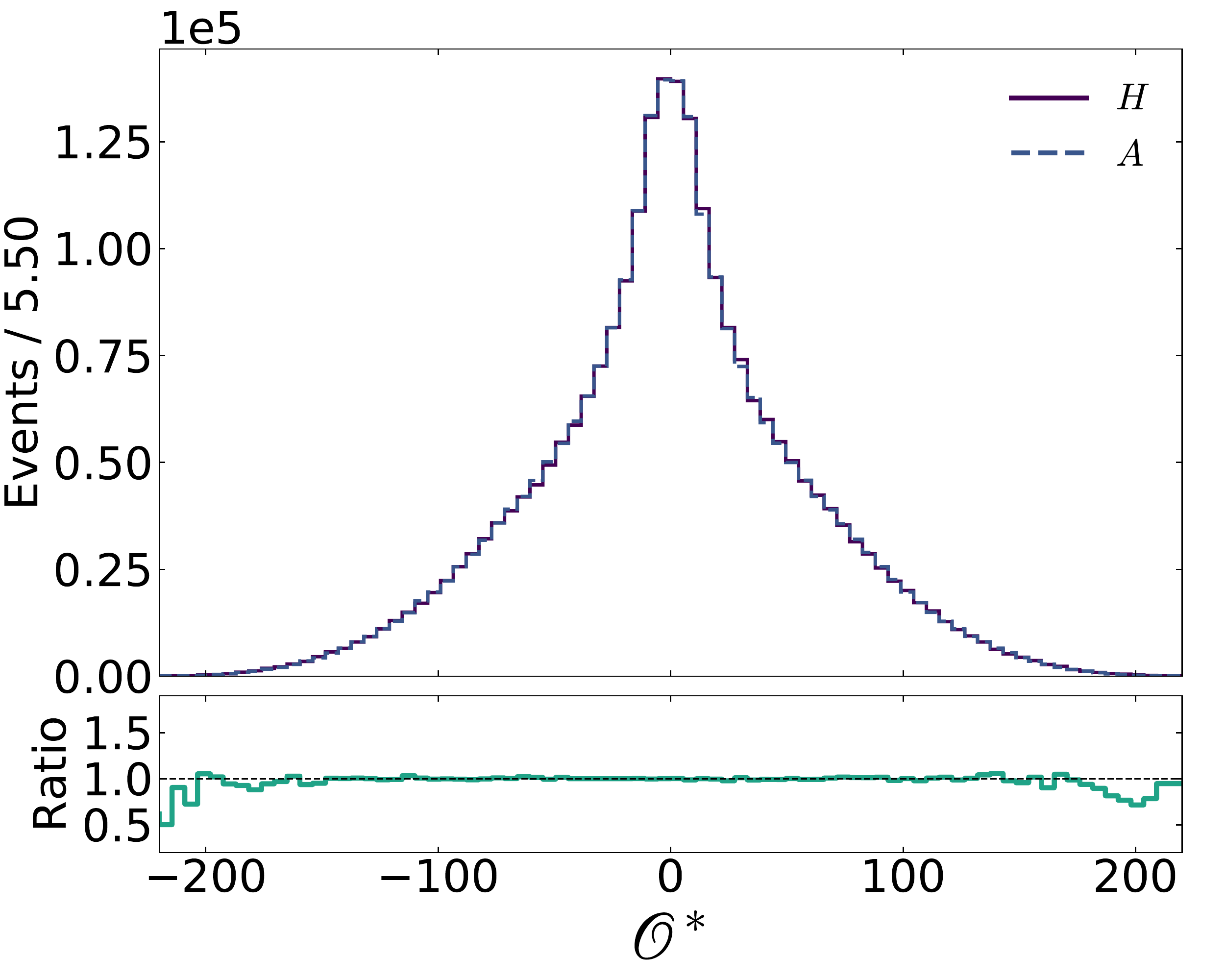}
		\label{fig:triple_corr}}
	\subfloat[][]{%
		\includegraphics[width=0.3\textwidth]{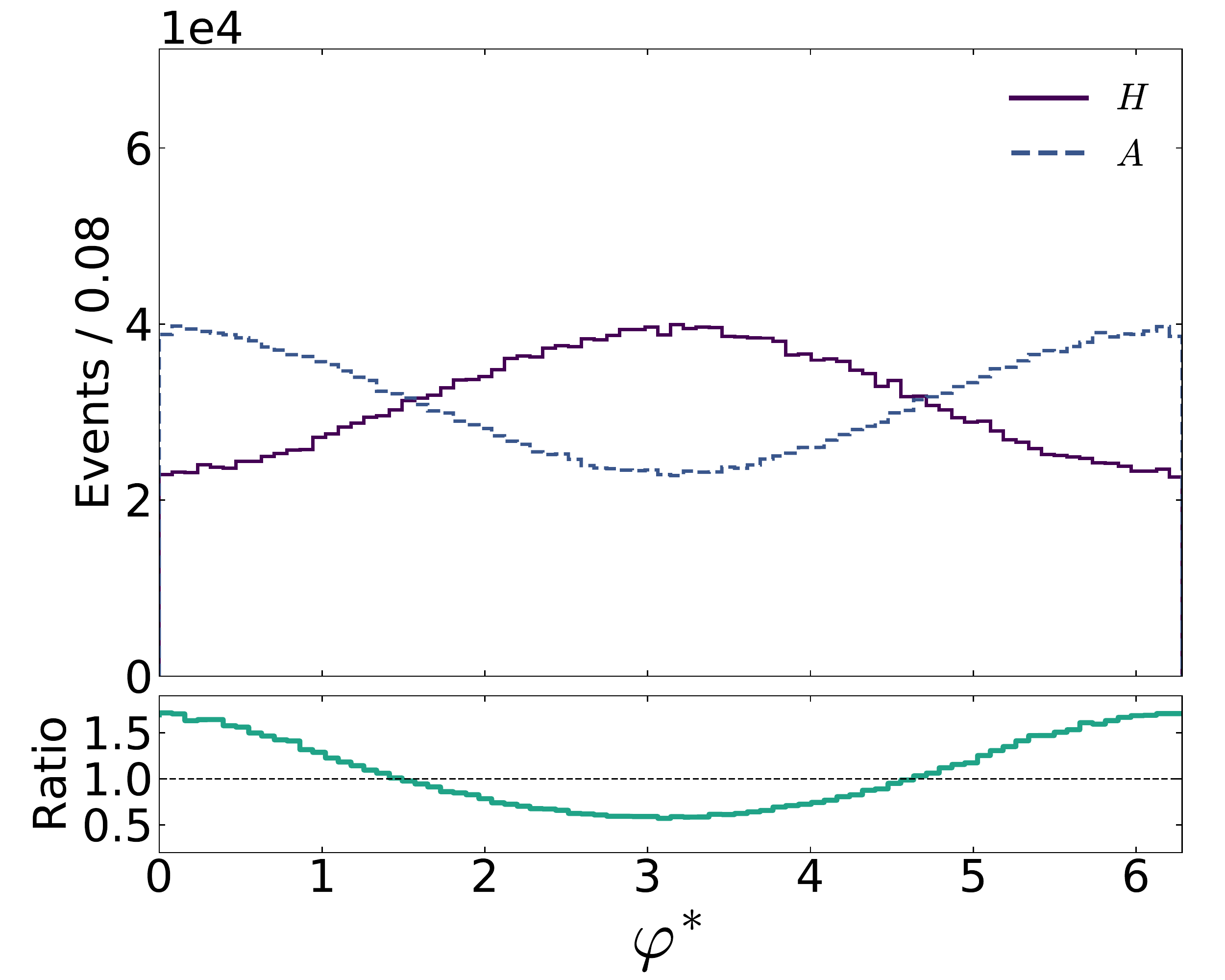}
		\label{fig:phistar}}
	\caption{\label{fig:features} Distributions for some kinematic features
        in $H \rightarrow \tau\tau \rightarrow (\pi^+\pi^0\nu)(\pi^-\pi^0\nu)$
        events (solid purple line) and $A \rightarrow \tau\tau \rightarrow
        (\pi^+\pi^0\nu)(\pi^-\pi^0\nu)$ events (dashed blue line), assuming
        $m_H=m_A=\unit[450]{GeV}$. The quantities in
        \protect\subref{fig:piminus_px} and \protect\subref{fig:pi0minus_pz}
        are momentum components of the $\pi^-$ and $\pi^0$ from the $\tau^-$
        decay, after each event has been boosted back to the visible ditau
        restframe and rotated such that the taus are back-to-back in the $z$
        direction and the $x$-component of the $\pi^+$ momentum is zero.  Panel
        \protect\subref{fig:transv_mass} shows the transverse mass
        $m_T^{\text{tot}}$, defined in~\cite{Patrignani:2016xqp}.  The
        observables $\Upsilon$~\protect\subref{fig:upsilon_minus} and
        $\scr{O}^{*}$~\protect\subref{fig:triple_corr}, defined in
        \cref{eq:upsilon} and \cref{eq:triplecorr}, respectively, are required
        for the computation of $\varphi^{*}$, along with the momentum vectors
        of the tau decay products. The distribution of $\varphi^{*}$ is shown
        in \protect\subref{fig:phistar}.  The green graph below each plot shows
        the ratio of the $A$-event and $H$-event distributions.
		} 
\end{figure*}
For the neural signal mixture estimation in~\cref{sec:network_method}, we train
a network to separate $H \to \tau\tau$ events from $A \to \tau\tau$ events.
The four-momenta of the visible tau decay products ($\pi^{\pm}$ and $\pi^{0}$)
constitute the most basic kinematic input features to our network.  The momenta
are boosted back to the visible ditau rest frame (the zero-momentum frame for
the four pions) and rotated so that the visible taus are back-to-back along the
$z$-axis. The system is then rotated a second time, now around the $z$-axis, so
that the $x$-component of the $\pi^+$ is zero.  This is done in order to align
all events to a common orientation, as the azimuthal angle around the $z$-axis
carries no physics information.

In addition to the pion momenta, the network is trained on missing transverse
energy ($E_T^{\text{miss}}$); the invariant mass of the four-pion system
($m_{\text{vis}}$); the transverse mass ($m_T^{\text{tot}}$); the impact
parameter vectors of the charged pions, which help constrain the neutrino
directions; the pion energy ratios $\Upsilon^{\pm}$, defined as
\begin{equation}
	\label{eq:upsilon}
	\Upsilon^{\pm} = \frac{E_{\pi^\pm}-E_{\pi^0}}{E_{\pi^\pm}+E_{\pi^0}}
	\, ;
\end{equation}
and the angle $\varphi^{*}$ between the tau decay planes.  For $\varphi^{*}$ we
follow the definition in~\cite{Berge:2015nua},\footnote{Our definition of
$\varphi^*$ only differs from that in~\cite{Berge:2015nua} in that we define
$\varphi^*$ in the $\pi^+\pi^0\pi^-\pi^0$ zero-momentum frame, whereas the
$\pi^+\pi^-$ zero-momentum frame is used in~\cite{Berge:2015nua}.} which uses
the direction $\hat{\mathbf{p}}_{\perp}^{(0)\pm}$ of the $\pi^{0}$ transverse
to the direction $\hat{\mathbf{p}}^{\pm}$ of the corresponding $\pi^{\pm}$, to
form an intermediate observable $\varphi \in [0, \pi)$ and a $CP$-odd triple
correlation product $\scr{O^{*}}$,
\begin{alignat}{2}	
    &\varphi &&= \arccos \big(\hat{\mathbf{p}}_{\perp}^{(0)+}
		\cdot \hat{\mathbf{p}}_{\perp}^{(0)-}\big) \quad
		\mathrm{and}\\ \label{eq:triplecorr}
    &\scr{O}^{*} &&= \hat{\mathbf{p}}^{+} \cdot
		\big(\hat{\mathbf{p}}_{\perp}^{(0)+}
		\times \hat{\mathbf{p}}_{\perp}^{(0)-}\big)\,.
\end{alignat}
From these, we can define an angle continuous on the interval $[0, 2\pi)$:
\begin{equation}
		\varphi' = 
		\begin{cases}
			\varphi & \mbox{if } \scr{O}^{*} \geq 0 \\
			2\pi - \varphi & \mbox{if } \scr{O}^{*} < 0 \\
		\end{cases}
		\, .
	\end{equation}
The distribution of $\varphi'$ depends on the sign of the product
$\Upsilon^{+}\Upsilon^{-}$; in the case of $\Upsilon^{+}\Upsilon^{-} \geq 0$,
the distribution is phase-shifted by $\pi$ relative to the case of
$\Upsilon^{+}\Upsilon^{-} < 0$. To incorporate this into a single consistent
$CP$-sensitive observable, we define $\varphi^{*}$ as
\begin{equation}
	\label{eq:phistar_definition}
		\varphi^{*} = 
		\begin{cases}
			\varphi' & \mbox{if } \Upsilon^{+}\Upsilon^{-} \geq 0 \\
			(\varphi' + \pi) \bmod 2\pi & \mbox{if } \Upsilon^{+}\Upsilon^{-} < 0 \\
		\end{cases}
		\, .
	\end{equation}

Before being input to the network, all feature distributions are standardised
to have zero mean and unit variance. A selection of the feature distributions
in the training data is shown in~\cref{fig:features}. The univariate feature
distributions are severely overlapping for $H$ and $A$ events, indicating that
the classification task is very challenging. The one feature which stands out
here is $\varphi^*$, which is the basis for the single-variable mixture
estimation described in~\cref{sec:phistar_method}. 

For the results presented in~\cref{sec:network_method} we use a network trained
on all features discussed above. However, features such as $\varphi^*$ and
$m_T$ are derived from the basic pion momenta that the network also has access
to.  These ``high-level'' features can in principle be inferred by the network
itself from the ``low-level'' pion momenta.  To briefly investigate this we
repeat the network training with varying subsets of the input features,
starting with only the pion four-momenta and sequentially adding $\varphi^*$,
$\Upsilon^{\pm}$ and the remaining features.  For all networks we obtain ROC
AUC scores of $\sim0.630$. While a full statistical comparison of the resulting
networks is beyond the scope of our study, this indicates that the network is
itself able to extract the relevant information from high-dimensional
correlations between the pion momenta, making the explicit inclusion of the
high-level inputs mostly redundant. We note that this observation is in
agreement with the results of~\cite{Baldi:2014kfa,Baldi:2014pta}.  

It is still interesting to investigate how much of the discriminatory power 
can be captured by the high-level features alone.
For this we train several classifiers on high-level
features only, adding a new set of features for each classifier.  The first
classifier is trained only on $\varphi^*$ and achieves a ROC AUC score of
$\sim0.605$. When $\Upsilon^{+}$ and $\Upsilon^{-}$ are included as input
features the performance improves to a score of $\sim0.618$.  This improvement
can be understood qualitatively from the fact that the difference between the
$\Upsilon^{\pm}$-conditional $\varphi^*$ distributions for $H$ and $A$ events
increases with $|\Upsilon^\pm|$. Adding $E_T^{\mathrm{miss}}$, $m_T$,
$\pi^\pm$ impact parameter vectors and $\scr{O^{*}}$ raises the ROC 
AUC score to $\sim0.620$, and finally including $m_{\text{vis}}$ 
further increases the score to $\sim0.623$, which seems to be the limit
for our network when trained on high-level features only.  This indicates that
$\varphi^*$ and $\Upsilon^{\pm}$ together capture most of the sensitivity, but
that the neural network is able to extract from the pion four-momenta some
additional information which is not contained in the high-level quantities.
Similar behaviour was seen in~\cite{Jozefowicz:2016kvz} in a study focusing on
the $CP$-nature of the $\unit[125]{GeV}$ Higgs.
%
\subsection{Network layout}
%
In this study we employ a fully-connected feed-forward network. The input layer
has 26 nodes, followed by 500 nodes in the first hidden layer, 1000 nodes in
the second hidden layer, and 100 nodes in the final hidden layer. These have
leaky ReLU~\cite{maas2013rectifier} activation functions, and
\\
dropout~\cite{srivastava2014dropout} is applied with a dropping probability of
0.375. No further regularisation is imposed. All network weights are
initialised from a normal distribution, following the He
procedure~\cite{HeZR015}. The output layer has a softmax activation function,
and we apply batch normalisation~\cite{IoffeS15} between all layers. The
weights are optimised using Adam~\cite{KingmaB14} with cross-entropy loss and
an initial learning rate of 0.03. 20\% of the training data are set aside to
validate the model performance during training. If there is no improvement of
the loss on the validation data for ten consecutive epochs, the learning rate
is reduced by a factor ten.  The network is trained for 100 epochs or until no
improvement is observed during 15 epochs, whichever occurs first.  The neural
network implementation is done using the Keras~\cite{chollet2015keras} and
TensorFlow~\cite{tensorflow2015-whitepaper} frameworks.
%
\section{The $\varphi^*$ method}
\label{sec:phistar_method}
%
\begin{figure}
\centering
	\subfloat[][]{%
		\includegraphics[width=0.4\textwidth]{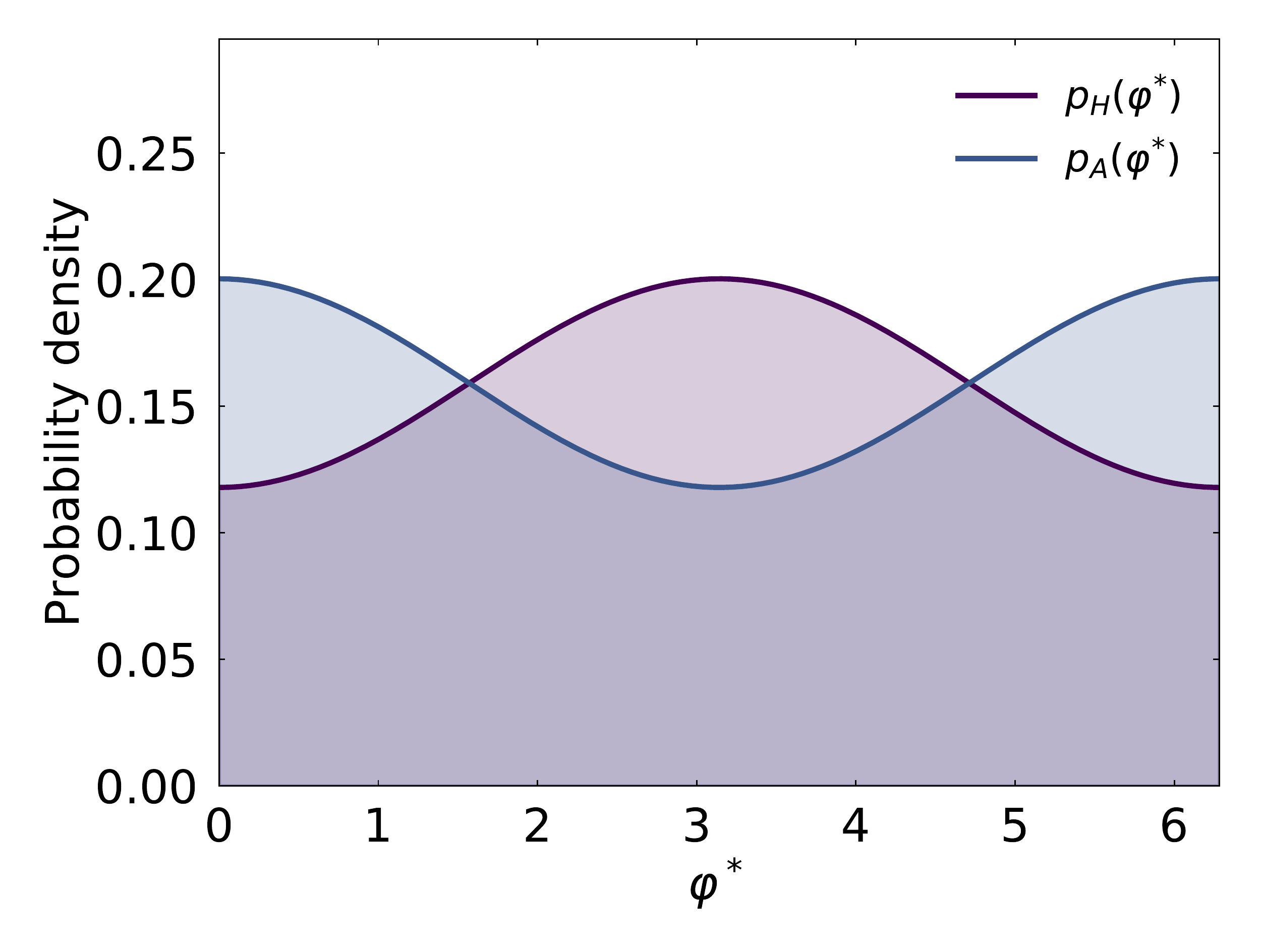}
		\label{fig:phistar_pdfs}
	}
	\\
	\subfloat[][]{%
		\includegraphics[width=0.4\textwidth]{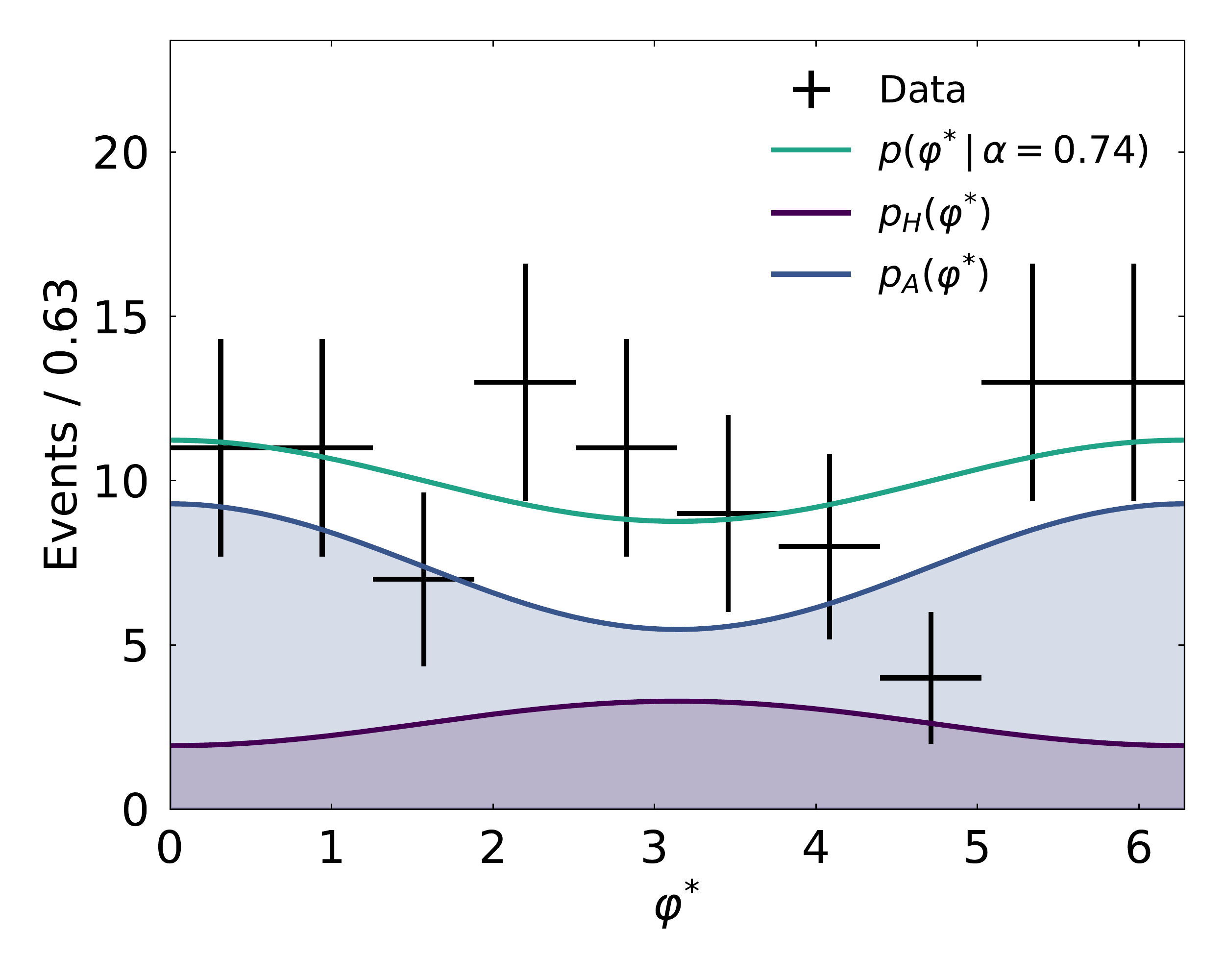}
		\label{fig:phistar_fit}
	}
	\caption{\label{fig:phistar_figs}\protect\subref{fig:phistar_pdfs} The
	probability density for $\varphi^*$ in $H$ events ($p_H(\varphi^*)$)
	and $A$ events ($p_A(\varphi^*)$).  \protect\subref{fig:phistar_fit} A
	fit of the mixture model $p(\varphi^*|\alpha) = \alpha p_A(\varphi^*) +
	(1-\alpha) p_H(\varphi^*)$ to a test dataset. Data points are shown in
	black, while the fitted model (normalized to 100 events) is shown in
	green.  For this dataset the best-fit $\alpha$ value is $\hat{\alpha} =
	0.74$.}
\end{figure}
Traditional approaches for separating $CP$-even and \mbox{-odd} decays are
based on the angle $\varphi^*$ between the tau decay planes, as defined
in~\cref{eq:phistar_definition}. The $\varphi^*$ distribution for $H$ and $A$
events can be seen in~\cref{fig:phistar_pdfs}.  The $CP$-sensitive parameter in
this distribution is the phase of the sinusoidal curve, which is shifted by
$\pi$ radians between the $H$ and $ A$ hypotheses. We note that the
distributions overlap across the full $\varphi^*$ range, hence no absolute
event separation is possible based on this variable. 

Using the simplified notation $p(\varphi^*|A) \equiv p_A(\varphi^*)$ and
$p(\varphi^*|H) \equiv p_H(\varphi^*)$, the $\varphi^*$ distribution for
$H$/$A$ signal data can be expressed as a simple mixture model,
\begin{equation}
	\begin{split}
		p(\varphi^*|\alpha) 
		&= \alpha p_A(\varphi^*) + (1-\alpha) p_H(\varphi^*)\\
		&= \alpha \big(a \cos\varphi^* + c\big) 
		+ \big(1-\alpha\big) \big(a \cos(\varphi^* + \pi) + c\big)\\
		&= \alpha\, a \cos\varphi^* + (1-\alpha)\, a \cos(\varphi^* + \pi) + c,
	\end{split}
\label{eq:phi_mixture_model}
\end{equation}
where we fix the amplitude $a$ and offset $c$ to $a=0.041$ and $c=0.159$,
obtained from a separate fit to $H$ and $A$ training data. This leaves us with
a model for the $\varphi^*$ distribution where $\alpha$ is the only free
parameter. Given a dataset $\{\varphi^*_i\}$ with $N$ events, we can now obtain
an estimate ${\hat{\alpha}}$ for $\alpha$ by maximising the likelihood function 
\begin{equation}
\begin{split}
    \hat{\alpha} &= \argmax \limits_{\alpha} \prod_{i=1}^{N}
	p(\varphi^*_i|\alpha)\,.\\ \tabularnewline
\end{split}
\end{equation}
We demonstrate this method in~\cref{fig:phistar_figs} for a dataset with 100
$H$/$A$ \textsf{Pythia} events, generated using a model with a true $\alpha$ of
$0.7$.  The pdfs $p_H(\varphi^*)$ and $p_A(\varphi^*)$ are shown
in~\cref{fig:phistar_pdfs}, while the fit result is shown
in~\cref{fig:phistar_fit}. For this example the best-fit $\alpha$ estimate
comes out at $\hat{\alpha} = 0.74$.

To demonstrate the statistical performance of this estimator we repeat the fit
using $10\,000$ independent test sets with $100$ \textsf{Pythia} events each,
generated with true $\alpha$ values of $0.5$, $0.7$ and $0.9$. The resulting
distributions of $\alpha$ estimates are shown in~\cref{fig:results_phistar},
where the purple (green) distributions depict results without (with) detector
effects.  By fitting a Gaussian to each distribution we find the spread in the
estimates to be $\sigma_\alpha = 0.27$ ($\sigma_\alpha^{\det} \simeq 0.45$)
when detector smearing is omitted (included).  Further, the estimator is
mean-unbiased for all three cases. Note that to demonstrate the unbiasedness we
have allowed the fit to vary $\alpha$ beyond the physically valid range of
$[0,1]$.
%
\section{The neural network method}
\label{sec:network_method}
%
When estimating some parameter $\theta$ using collider data we ideally want to
make use of the multivariate density $p(\x|\theta)$ for the complete set of
event features $\x$.\footnote{Here $\theta$ represents an arbitrary model
parameter, not necessarily a simple mixture parameter.} However, it is
typically infeasible to evaluate this density directly for a given $\x$.  A
common approach is then to construct a new variable $y(\x)$ and base the
parameter estimation on the simpler, univariate distribution $p(y(\x)|\theta)$,
as exemplified by the $\varphi^*$ fit in~\cref{sec:phistar_method}. 

The performance of such a univariate approach depends on how well the
distribution $p(y(\x)|\theta)$ retains the sensitivity to $\theta$ found in the
underlying distribution $p(\x|\theta)$. In the special case where the map
$y(\x)$ is the output from a trained classifier, it can be shown that using
$p(y(\x)|\theta)$ to estimate $\theta$ in the ideal limit is equivalent to
using the full data distribution $p(\x|\theta)$. Here we briefly review this
argument before applying the classifier approach to our mixture estimation
problem.

After training on $\theta$-labeled data, a classifier that minimizes a suitably
chosen error function will approximate a decision function $s(\x)$ that is a
strictly monotonic function of the density ratio
$p(\x|\theta)/p(\x|\theta')$~\cite{friedman2000}.\footnote{In general the
decision function can depend directly on the parameter values $\theta$ and
$\theta'$: $s = s(\x;\theta,\theta')$. However, this is not the case for a
mixture estimation problem like the one considered here, where $\x$ represents
a single draw from one of the mixture model components (kinematic data from a
single $H$ or $A$ event) and the parameter of interest is the unknown component
mixture ($\alpha$) of the complete dataset $\{\x_i\}$.} 
As shown in~\cite{Cranmer:2015bka}, the monotonicity of $s(\x)$ ensures that
density ratios based on the multivariate distribution $p(\x|\theta)$ and the
univariate distribution $p(s(\x)|\theta)$ are equivalent,
\begin{equation}
	\frac{p(\x|\theta)}{p(\x|\theta')}
	= \frac{p(s(\x)|\theta)}{p(s(\x)|\theta')}.
\end{equation}
If we now take $\theta'$ to be a fixed value such that the support of
$p(\x|\theta')$ covers the support of $p(\x|\theta)$,\footnote{This is 
trivially satisfied for any choice $\theta' \in (0,1)$ when $\theta$ 
represents the mixture parameter of a simple two-component mixture model.} 
the maximum likelihood estimator for $\theta$ based on $p(\x|\theta)$ 
can be rewritten as follows~\cite{Cranmer:2015bka}:
\begin{equation}
\begin{split}
	\hat{\theta} &= \argmax \limits_{\theta} \prod\limits_{i=1}^N p(\x_i|\theta)\\
	&= \argmax \limits_{\theta} \prod\limits_{i=1}^N \frac{p(\x_i|\theta)}{p(\x_i|\theta')}\\
	&= \argmax \limits_{\theta} \prod\limits_{i=1}^N
	\frac{p(s(\x_i)|\theta)}{p(s(\x_i)|\theta')}\\
	&= \argmax \limits_{\theta} \prod\limits_{i=1}^N p(s(\x_i)|\theta).
\end{split}
\label{eq:MLE_ratio}
\end{equation}
Hence, if the classifier output $y(\x)$ provides a reasonable approximation of
$s(\x)$ 
we can expect the maximum likelihood estimator based on $p(y(\x)|\theta)$
to exhibit similar performance to an estimator based on $p(\x|\theta)$.  
%
%
The main drawbacks of this approach are the complications associated with
training the classifier, and that the physics underlying the parameter
sensitivity may remain hidden from view.

\begin{figure}
\centering
	\subfloat[][]{%
		\includegraphics[width=0.4\textwidth]{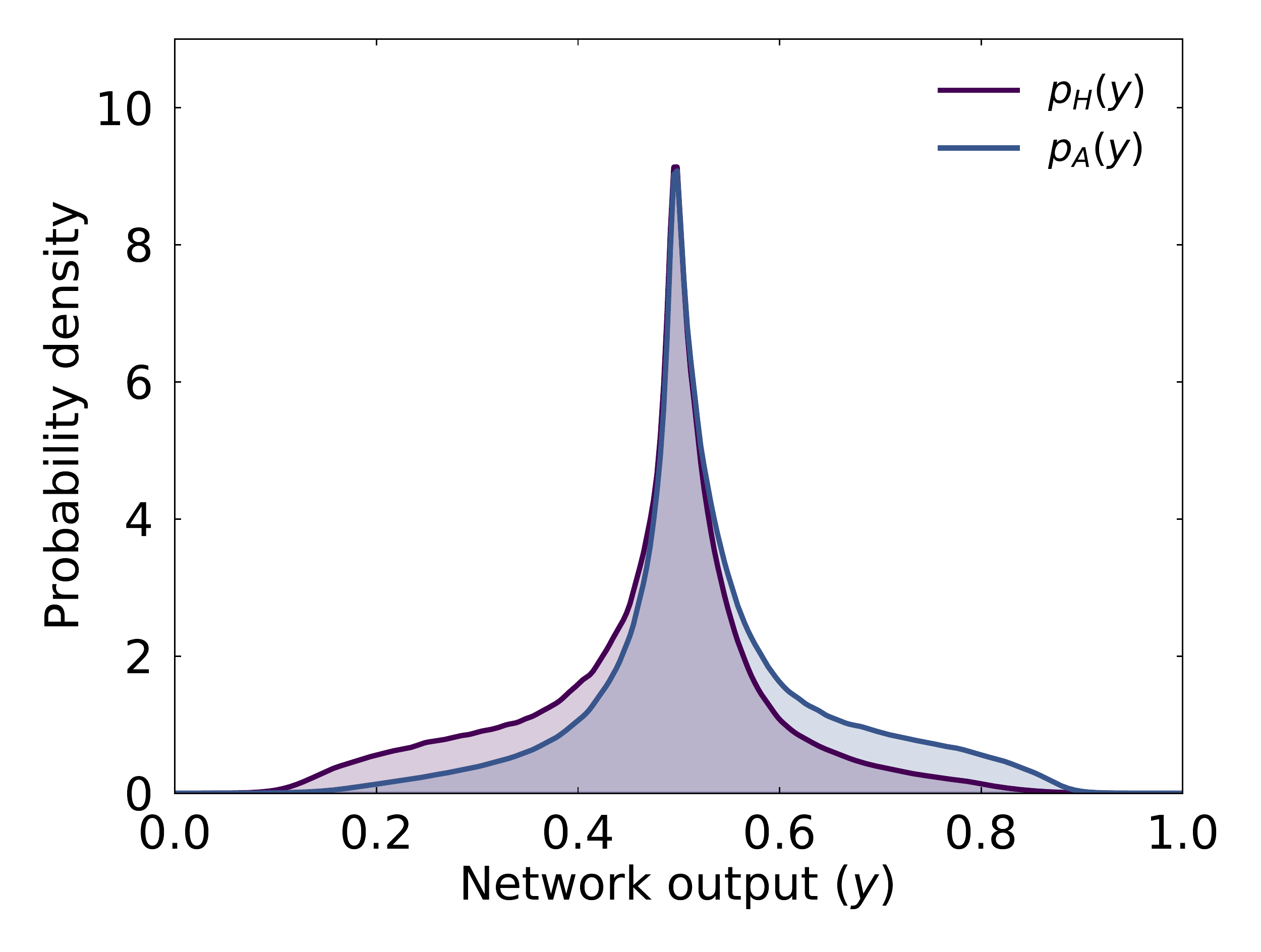}\label{fig:mle_nn_pdfs}
	}
	\\
	\subfloat[][]{%
		\includegraphics[width=0.4\textwidth]{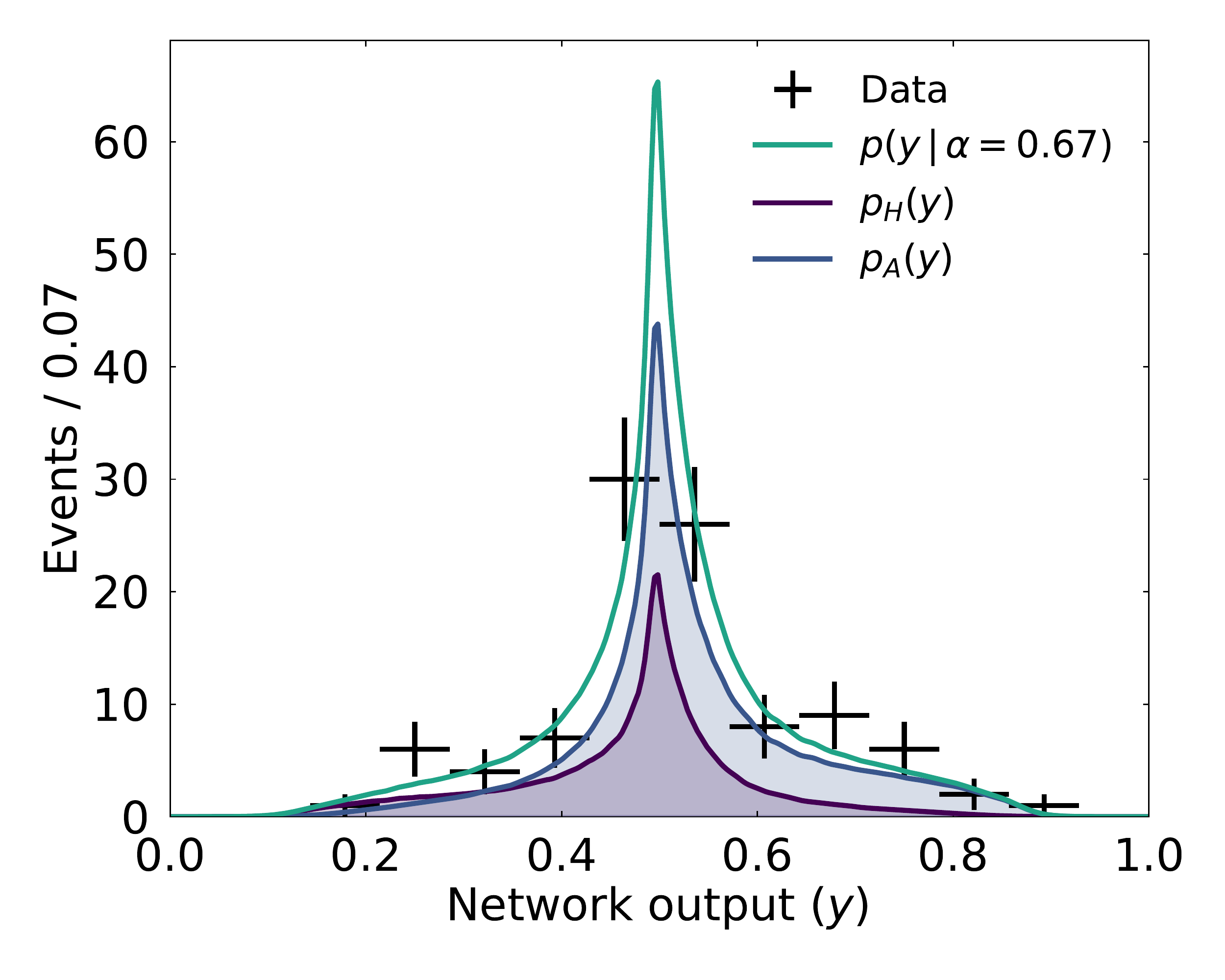}\label{fig:mle_nn_fit}
	}	
	\caption{\protect\subref{fig:mle_nn_pdfs} KDE estimate for the
	distribution of the network output $y$ for $H$ events ($p_H(y)$) and
	$A$ events ($p_A(y)$), given a balanced network.
	\protect\subref{fig:mle_nn_fit} A fit of the mixture model $p(y|\alpha)
	= \alpha p_A(y) +  (1-\alpha) p_H(y)$ to the same example dataset as
	used in~\cref{fig:phistar_fit}. Data points are shown in black and the
	fitted model in green. The best-fit $\alpha$ value is $\hat{\alpha} =
	0.67$.}
\end{figure}

We now apply this classifier approach to our $H$/$A$ mixture estimation
problem.  The maximum likelihood estimator for the mixture parameter $\alpha$
is then given by 
\begin{equation}
	\label{eq:model_y}
	%
	\begin{split}
		\hat{\alpha} &= \argmax \limits_{\alpha} \prod\limits_{i=1}^N
		p(y(\x_i)|\alpha)\\
		&= \argmax \limits_{\alpha} \prod\limits_{i=1}^N \Big[
		\alpha p_A(y(\x_i)) +  (1-\alpha) p_H(y(\x_i))\Big],\! 
	\end{split}
\end{equation}
where we have expressed the overall network output distribution $p(y|\alpha)$
as a mixture of the pure-class distributions $p(y|A) \equiv p_A(y)$ and $p(y|H)
\equiv p_H(y)$.  We use a network trained on a balanced set of $H$ and $A$
events. The network is trained to associate outputs $y=0$ and $y=1$ with $H$
and $A$ events, respectively.  By applying this network to another labeled
dataset of equal size to the training set, we construct templates for the
probability densities $p_H(y)$ and $p_A(y)$ in \cref{eq:model_y} using a
nonparametric kernel density estimation method (KDE)~\cite{Cranmer:2000du}.
The resulting templates are shown in \cref{fig:mle_nn_pdfs}.  We note that the
pdfs do not span the entire allowed range $y \in [0,1]$. This is expected,
since the $CP$ nature of a single event cannot be determined with complete
certainty.  Proper determination of the pdf shapes in the extremities --- where
the sensitivity is highest --- requires a sufficient amount of data, which is
why we devote a similarly sized data set to the template creation as to the
network training.

Given a set of unlabeled data we can now estimate $\alpha$ by carrying out the
maximization in \cref{eq:model_y} as an unbinned maximum-likelihood fit. The
resulting fit to the same example dataset as used for the $\varphi^*$ fit
in~\cref{fig:phistar_fit} is shown in \cref{fig:mle_nn_fit}.  The best-fit
$\alpha$ estimate in this case is $\hat{\alpha} = 0.67$. 
%
\subsection{Results}
\label{subsec:results}
\begin{figure*}
\centering
	\subfloat[][]{%
	\includegraphics[width=0.4\textwidth]{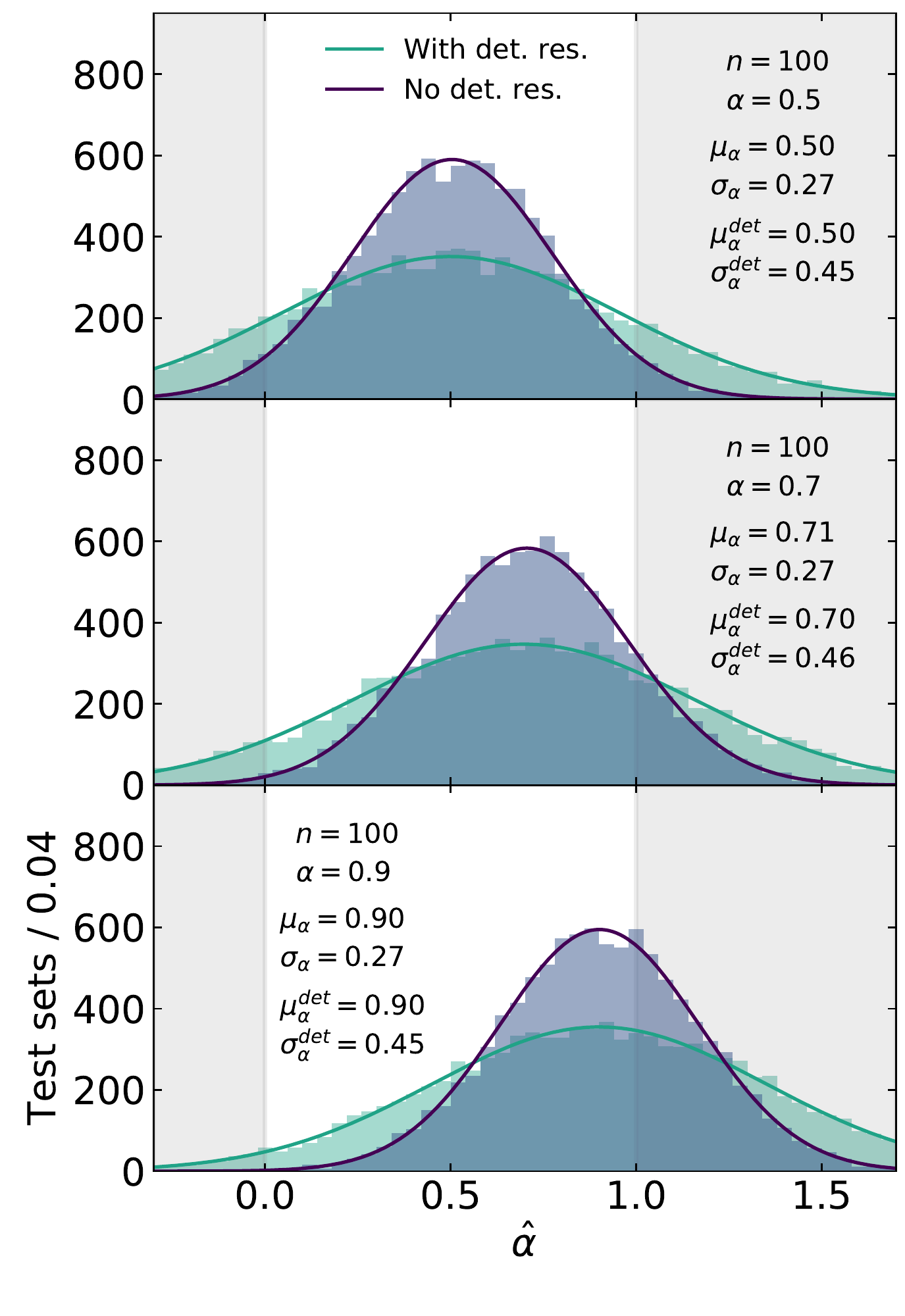}\label{fig:results_phistar}
	}
	\subfloat[][]%
	{\includegraphics[width=0.4\textwidth]{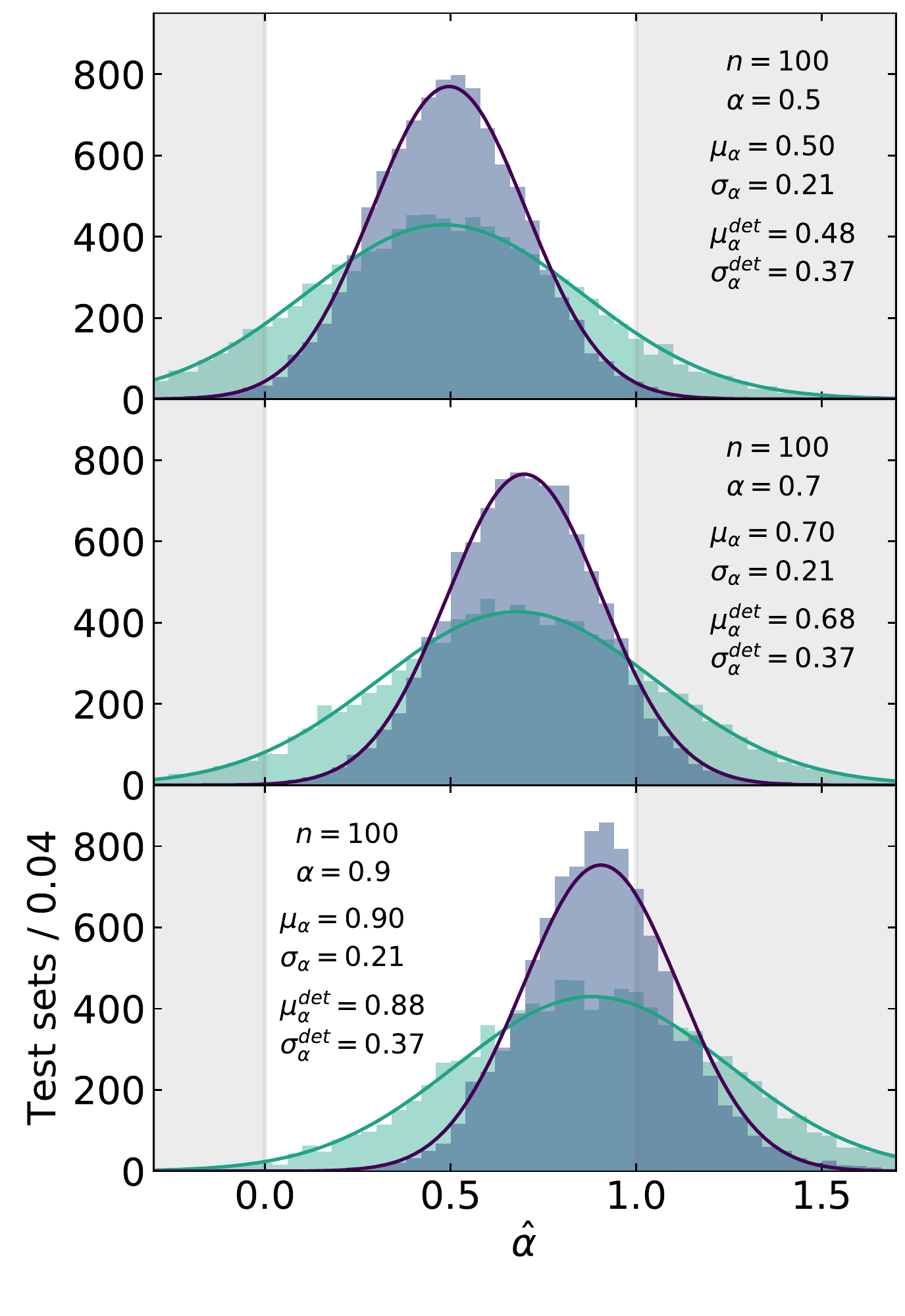}\label{fig:results_nn}
	}
	\caption{\label{fig:results}Comparison of the distributions of $\alpha$
	estimates using~\protect\subref{fig:results_phistar} the $\varphi^{*}$
	method and~\protect\subref{fig:results_nn} the neural network method,
	for test sets generated with $\alpha = 0.5$ (top), $\alpha = 0.7$
	(middle) and $\alpha = 0.9$ (bottom).  The slight deviation from
	Gaussianity seen around $\hat{\alpha} = 1.2$ in the bottom right plot
	is due to the fact that we let $\alpha$ vary beyond $[0,1]$ in our
	fits, but still demand that that the mixture model $p(y|\alpha)$ is
	always non-negative. See the text for further details.  }
\end{figure*}

%
We can now compare the performance of the neural network method with that of
the $\varphi^*$ method of~\cref{sec:phistar_method}. To this end, we apply the
network method to the same test sets as used in~\cref{fig:results_phistar},
\ie~$10\,000$ datasets of 100 \textsf{Pythia} events each, for each of the
three scenarios \mbox{$\alpha$ = 0.5, 0.7, 0.9}. The analysis is repeated with
network training and test sets  with and without detector smearing.  The
results are given in~\cref{fig:results_nn}, for easy comparison with the
corresponding results of the $\varphi^*$ method in~\cref{fig:results_phistar}.
We fit each distribution of $\alpha$ estimates with a Gaussian and summarize
the fit parameters in~\cref{tab:results}.

As for the $\varphi^*$ method, we find that detector smearing significantly
impacts the width of the $\alpha$ distribution, which increases from
$\sigma_\alpha = 0.21$ to $\sigma_\alpha^{\det} = 0.37$ upon inclusion of
detector effects.  Yet, the network approach consistently outperforms the
$\varphi^*$ method, as $\sigma_\alpha$ and $\sigma_\alpha^{\det}$ are reduced by
$\sim22$\% and $\sim18$\%, respectively, compared to the $\varphi^*$ results.
So while the absolute widths of the $\alpha$ distributions
in~\cref{fig:results} illustrate that a dataset of 100 events is probably too
small to obtain an accurate $\alpha$ estimate, the comparison with the
$\varphi^*$ results indicates that the relative performance gain offered by the
network approach is relatively robust against detector smearing.

Similar to the $\varphi^*$ method, the network method provides a mean-unbiased 
estimator. In order to demonstrate this we allow $\alpha$ to
vary outside the physical range $[0,1]$ in our fits.  However, for $\alpha >
1$, the combined mixture model $p(y|\alpha) = \alpha p_A(y) +  (1-\alpha)
p_H(y)$ will become negative for $y$ values that satisfy $p_A(y)/p_H(y) <
(\alpha-1)/\alpha$.  This we do not allow in our fits, and in such cases we
lower the $\alpha$ estimate until $p(y|\alpha)$ is non-negative everywhere.
This choice explains the slight deviation from Gaussianity in the region around
$\hat\alpha = 1.2$ in the bottom right plot.\footnote{The same effect is not
seen for the $\varphi^*$ fits, as the ratio $p_A(\varphi^*)/p_H(\varphi^*) \geq
0.59$ for all $\varphi^*$, and none of the test sets prefer an $\alpha$ value
as large as $1/(1-0.59)\approx2.4$.}

\begin{figure*}
\centering
	\subfloat[][]{%
	\includegraphics[width=0.4\textwidth]{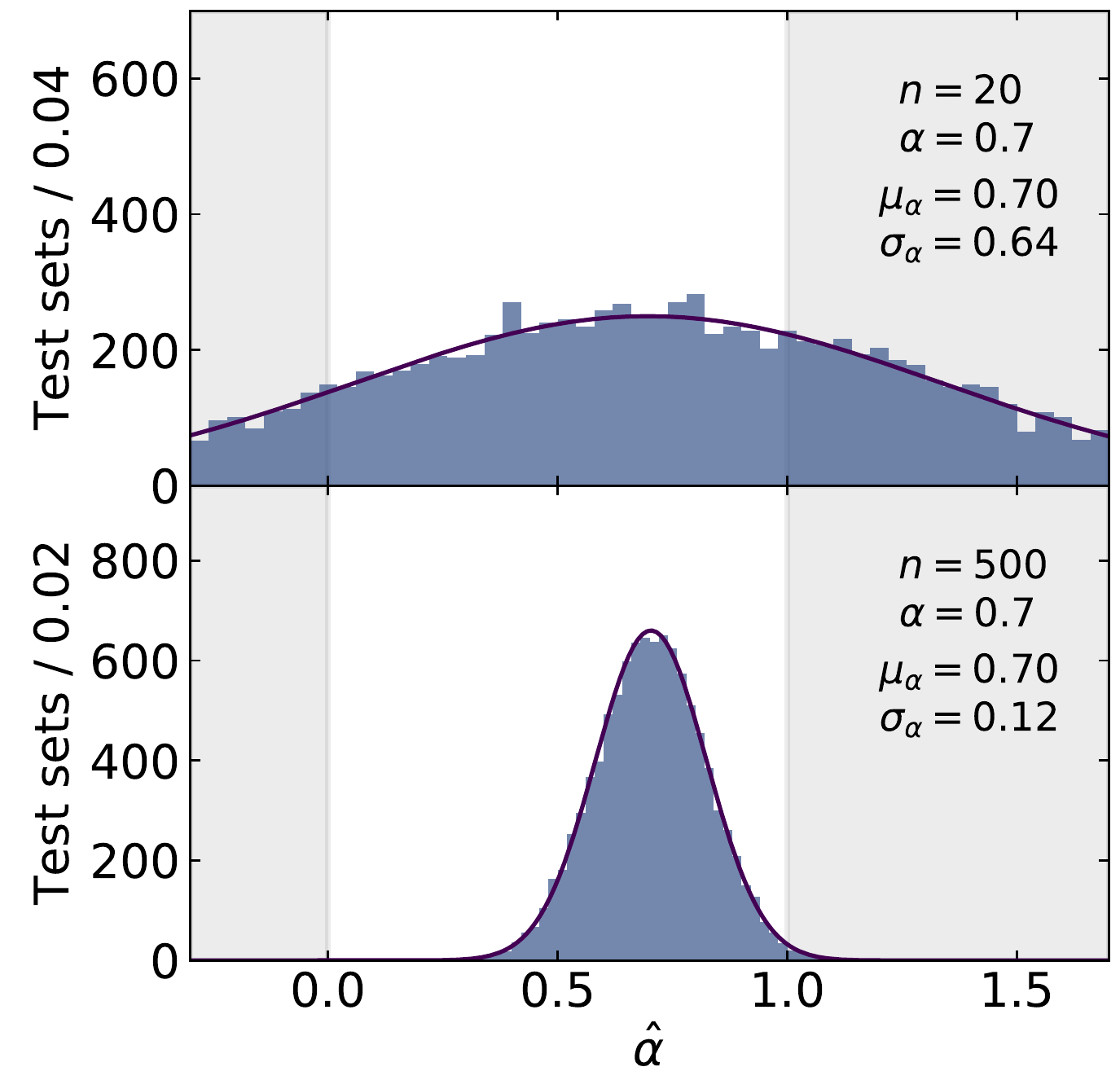}
	\label{fig:results_20_vs_500_phistar}
	}
	\subfloat[][]{%
	\includegraphics[width=0.4\textwidth]{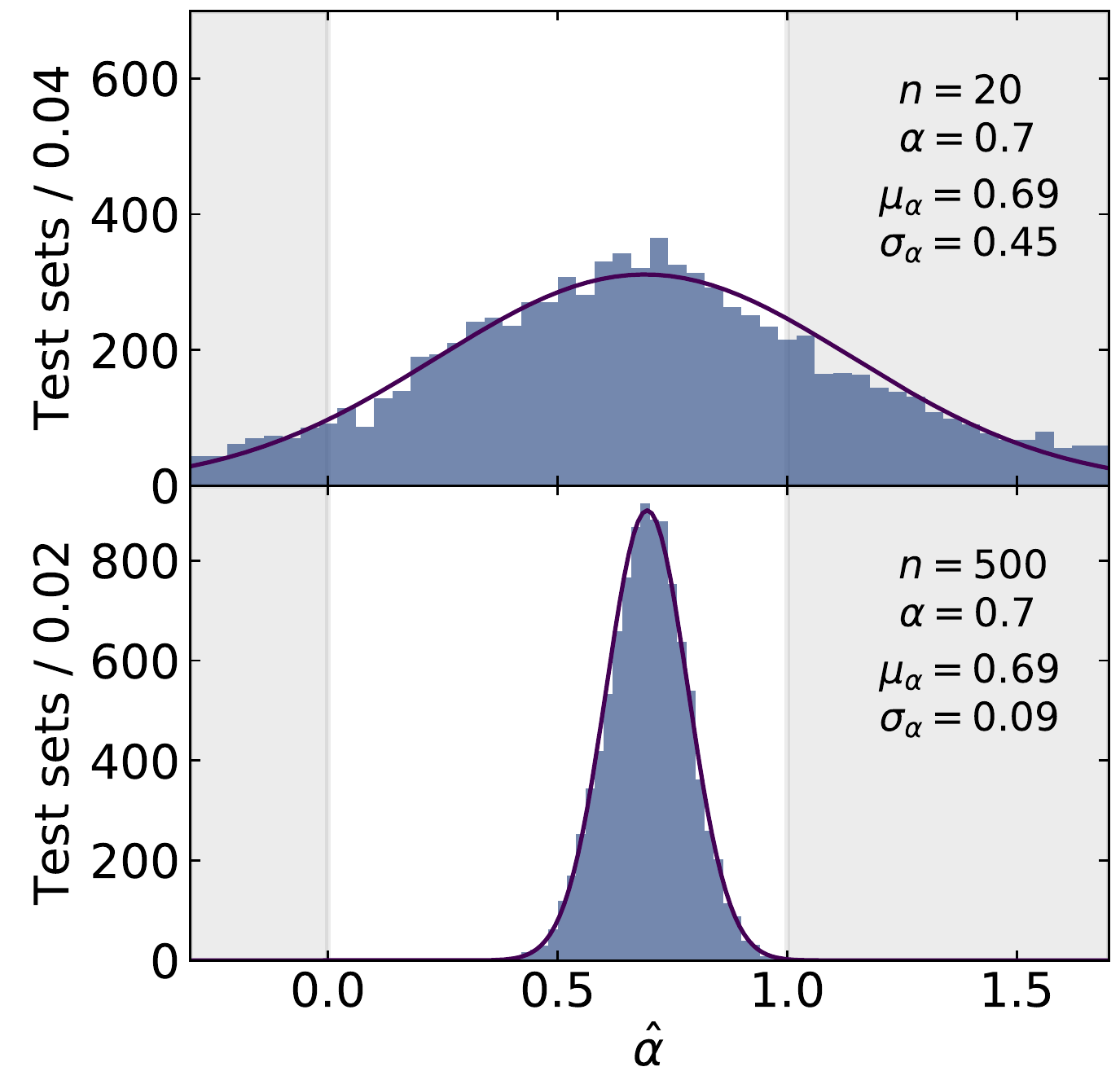}
	\label{fig:results_20_vs_500_mle}
	}
	\caption{\label{fig:results_20_vs_500}Comparison of the distributions
	of $\alpha$ estimates
	using~\protect\subref{fig:results_20_vs_500_phistar} the $\varphi^{*}$
	method and~\protect\subref{fig:results_20_vs_500_mle} the neural
	network method, for test sets generated with $\alpha = 0.7$. The top
	row shows results for test sets containing 20 events each, while
	the bottom row corresponds to test sets with 500 events each. The
	deviation from the Gaussian distribution seen at high $\alpha$ in the
	upper right plot is due to the same effect as discussed
	for~\cref{fig:results}.  }
\end{figure*}

\Cref{fig:results_20_vs_500} shows the distributions of $\alpha$ estimates for
the cases of 20 events per test dataset (top row) and 500 events per test
dataset (bottom row), where all sets have been generated with $\alpha = 0.7$
and no detector smearing has been included. Compared to the results with 100
events per set, $\sigma_\alpha$ for both fit methods increase (decrease) by
approximately a factor $\sqrt{5}$ for the case with 20 (500) events per set, as
expected from the factor 5 decrease (increase) in statistics. Thus, the
relative accuracy improvement of the neural network approach over the
$\varphi^{*}$ method remains approximately the same: 30\% for the 20-events
case, and 25\% for the 500-events case. However, the absolute spread of
estimates in the 20-events case shows that this is clearly not enough
statistics to obtain a useful estimate of $\alpha$.

\begin{figure}
\centering
	\subfloat[][]{%
	\includegraphics[width=0.4\textwidth]{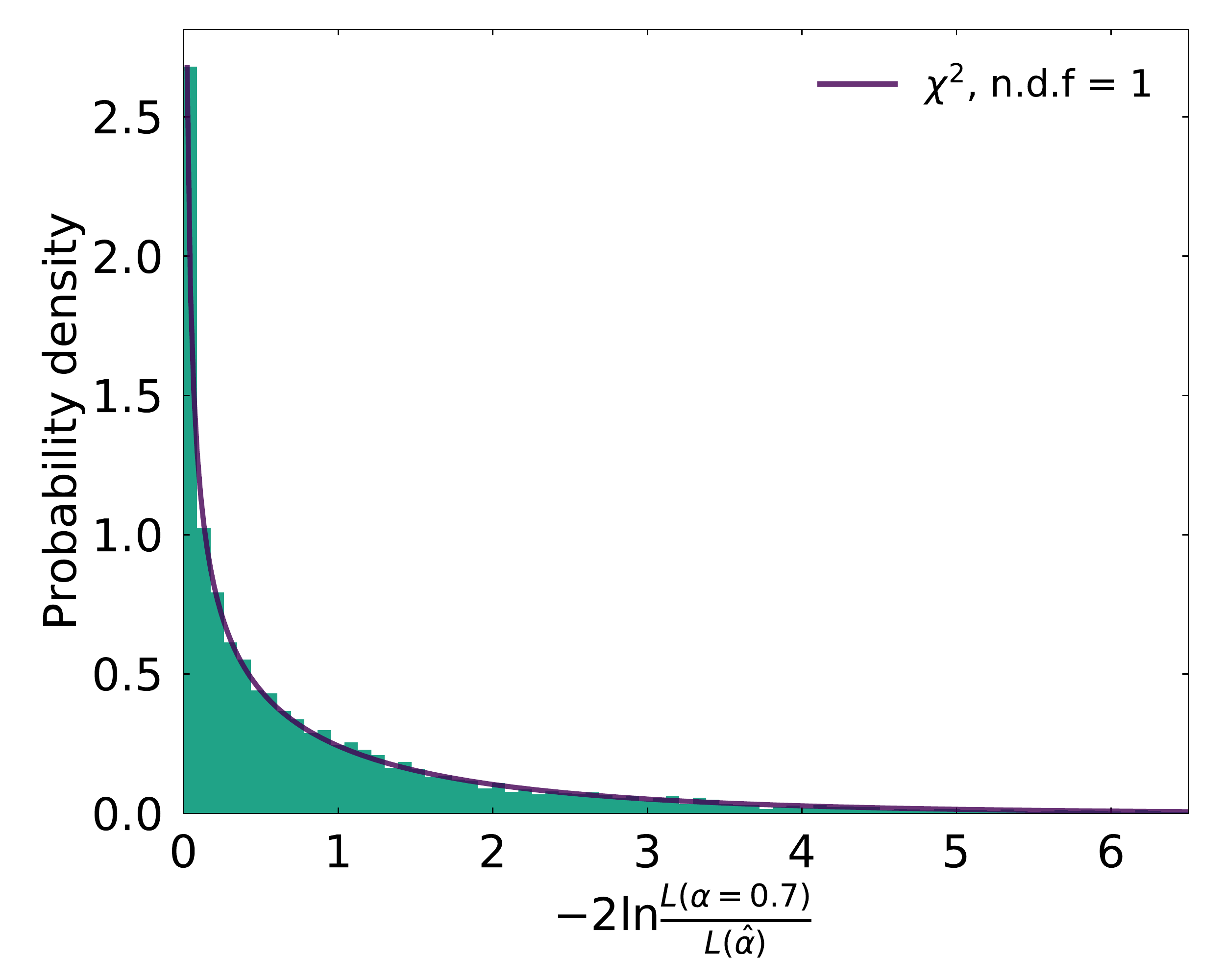}
	\label{fig:loglambda}
	}
	\\
	\subfloat[][]{%
	\includegraphics[width=0.4\textwidth]{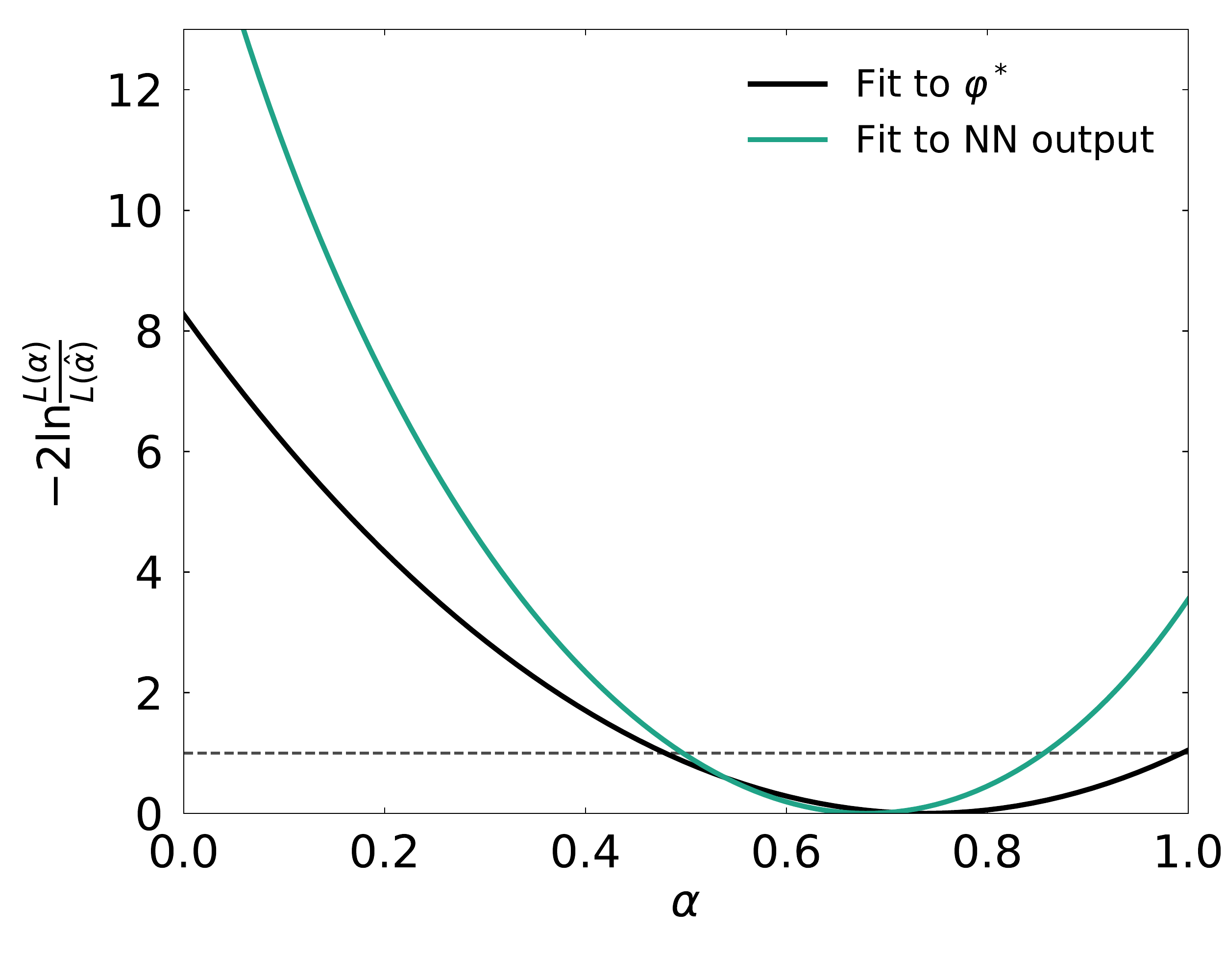}
	\label{fig:nll_comparison}
	}
	\caption{\label{fig:comparisons}\protect\subref{fig:loglambda}
	Distribution of the log-likelihood ratio
	$-2\ln(L(\alpha=0.7)/L(\hat\alpha))$ for the $10\,000$ test sets
	generated with $\alpha = 0.7$. Overlaid is a $\chi^2$ distribution for
	one degree of freedom. \protect\subref{fig:nll_comparison} Comparison
	of the log-likelihood ratio curves for the test dataset
	from~\cref{fig:phistar_fit,fig:mle_nn_fit}, using the network method
	(green) and the $\varphi^*$ method (black). Intersection with the
	horizontal dashed line at $-2\ln(L(\alpha)/L(\hat{\alpha})) = 1$
	illustrates the $1\sigma$ confidence intervals, which for this example
	are $[0.48,1.0]$ for the $\varphi^{*}$ method and $[0.50,0.86]$ for the
	neural network method.
    }
\end{figure}

As a cross-check of the behaviour of the network fit method, we plot
in~\cref{fig:loglambda} the distribution of the log-likelihood ratio
$-2\ln(L(\alpha=0.7)/L(\hat\alpha))$ for all test datasets of our benchmark
point with $\alpha = 0.7$.  According to Wilks' theorem~\cite{Wilks:1938dza},
the distribution of this statistic should tend towards a $\chi^2$ distribution
with one degree of freedom.  By overlaying a $\chi^{2}$ distribution
in~\cref{fig:loglambda} we see that this is indeed the case.  Thus, confidence
intervals constructed from the log-likelihood ratio for a neural network fit
should have the expected coverage.  
In~\cref{fig:nll_comparison} we show the log-likelihood ratio curves for the
example dataset used in~\cref{fig:phistar_fit,fig:mle_nn_fit}. The narrowing of
the log-likelihood parabola for the network method again illustrates the
increase in precision over the $\varphi^*$ method.
\begin{table*}
	\centering
	\caption{\label{tab:results}Summary of $\alpha$ estimation on $10\,000$
	independent test sets with 100 events in each set, using the
	$\varphi^*$ fit and the neural network (NN) fit methods.} 
	\begin{tabular}{l c c c}
		\toprule
		True mixture parameter $\alpha$ 
		& $\alpha=0.5$ & $\alpha=0.7$ & $\alpha=0.9$ \\ 
		\midrule
		$\alpha$ estimates ($\varphi^*$ method, no detector smearing)    & $0.50 \pm 0.27$ & $0.71 \pm 0.27$ & $0.90 \pm 0.27$ \\
		$\alpha$ estimates ($\varphi^*$ method, with detector smearing)  & $0.50 \pm 0.45$ & $0.70 \pm 0.46$ & $0.90 \pm 0.45$ \\
		$\alpha$ estimates (NN method, no detector smearing)        & $0.50 \pm 0.21$ & $0.70 \pm 0.21$ & $0.90 \pm 0.21$ \\
		$\alpha$ estimates (NN method, with detector smearing)      & $0.48 \pm 0.37$ & $0.68 \pm 0.37$ & $0.88 \pm 0.37$ \\
		\bottomrule
	\end{tabular}
\end{table*}

For this study we focus only on the separation of two signal classes, not the
separation of signal from background.  Of course, a realistic dataset is likely
to contain a significant fraction of background events. For the signal scenario
studied here, the most important backgrounds are due to ``fake taus'' from QCD
production, single $Z$ production ($pp\to Z\to\tau\tau$), double $Z$ and $W$
production ($pp\to ZZ/WZ/WW\to\tau\tau+X$) and top pair production
($t\bar{t}\to WbWb \to \tau\tau+X$).  While such backgrounds will degrade the
absolute accuracy in the signal mixture estimate, it is likely to impact the
$\varphi^*$ method more severely than the neural network method.  With one or
several background components in the mixture model, the network's ability to
extract information from the many-dimensional kinematic space should allow it
to differentiate the background components from the signal components better
than what is possible with the $\varphi^*$ variable alone. We therefore expect
a similar or better relative performance of the network method in the presence
of background, compared to the results we have presented here.  There are two
ways to extend the network method to take into account additional components in
the mixture model: either by implementing a multi-class classifier, or by
training multiple binary classifiers on pairwise combinations of the model
components. Based on~\cite{Cranmer:2016swd} we expect the latter approach would
give the best performance.
%
\section{Conclusions}
\label{sec:conclusion}
%
Estimating the component weights in mixture models with largely overlapping
kinematics is a generic problem in high-energy physics. In this paper we have
investigated how a deep neural network approach can improve signal mixture
estimates in the challenging scenario of a ditau LHC signal coming from a pair
of heavy, degenerate Higgs bosons of opposite $CP$ charge. This is a 
theoretically well-motivated scenario within both general and more constrained
 Two-Higgs-Doublet Models. 

We have studied a benchmark scenario with degenerate $H$ and $A$ states at $m_H
= m_A = \unit[450]{GeV}$.  For this case we find that the neural network
approach provides a $\sim20\%$ reduction in the uncertainty of signal mixture
estimates, compared to estimates based on fitting the single most
discriminating kinematic variable ($\varphi^*$). However, the improved accuracy
of the neural network approach comes with a greater computational complexity.

The network method we have studied here can be extended to include additional
mixture components, such as one or several background processes, either by
training a multi-class classifier or by training multiple binary classifiers.
To increase the available statistics, the method can also be extended to work
with a wider range of tau decay modes, for instance by using the ``impact
parameter method'' described in~\cite{Berge:2015nua}.

The code used to generate events, train the network and run the maximum
likelihood estimates will be made available on \url{gitlab.com/BSML} after
publication. 
%
\section*{Acknowledgements}
%
We would like to thank Andrey Ustyuzhanin and Maxim Borisyak for helpful
discussions during MLHEP 2017, and Kyle Cranmer for comments and discussions
during Spåtind 2018. We also thank Ørjan Dale for helpful comments. S.M.\ and
I.S.\ thank the Theory Section at the Department of Physics, University of
Oslo, for the kind hospitality during the completion of this work. This work was
supported by the Research Council of Norway through the FRIPRO grant 230546/F20.
%
%
\begin{appendices}
\crefalias{section}{appsec}
\section{Supplementary figures}
\label{sec:supplementary_figures}
\begin{figure*}
\centering
	\subfloat[][]{%
	\includegraphics[width=0.3\textwidth]{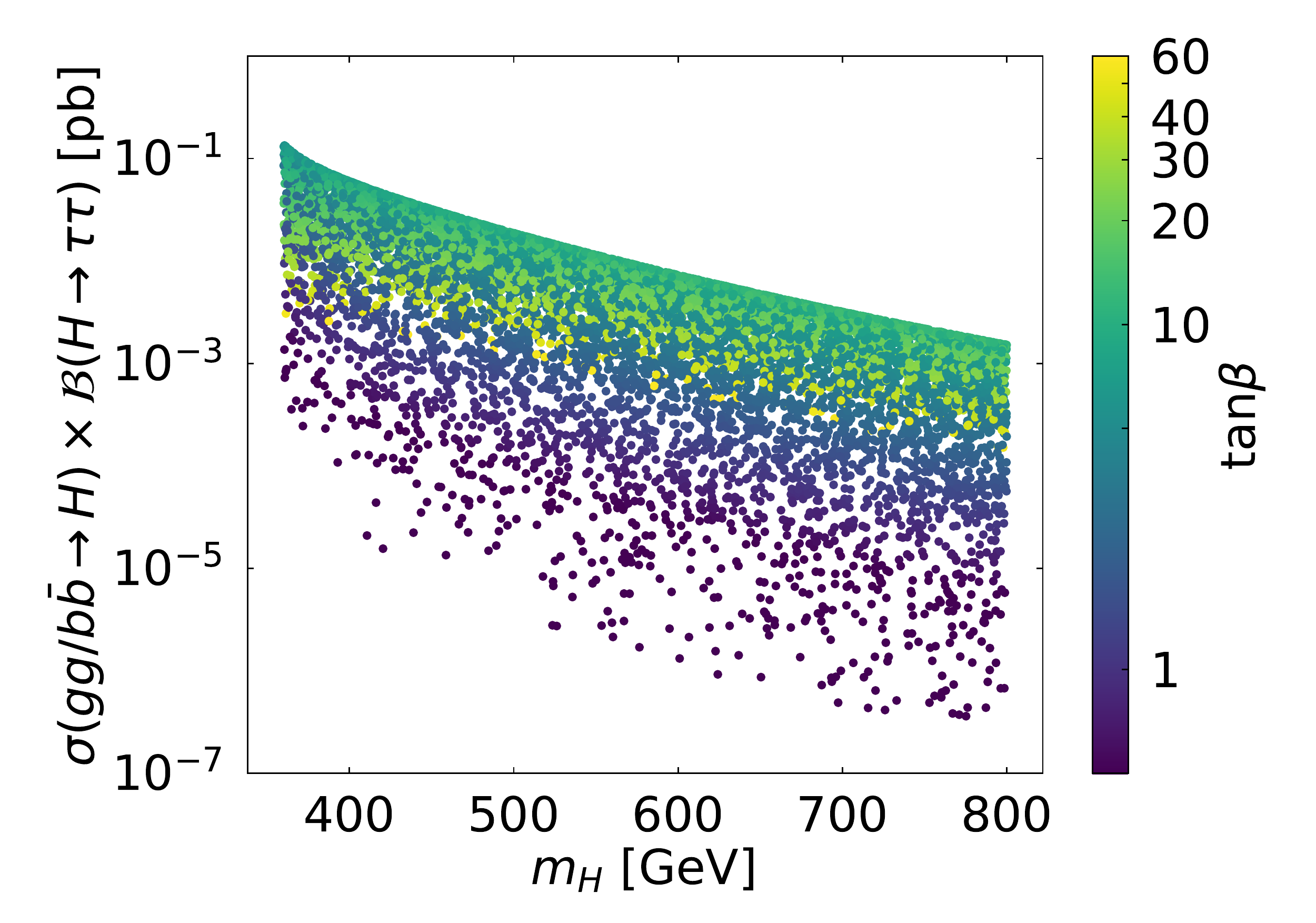}
	\label{fig:xsecscanH_typeIV}
	}
	\subfloat[][]{%
	\includegraphics[width=0.3\textwidth]{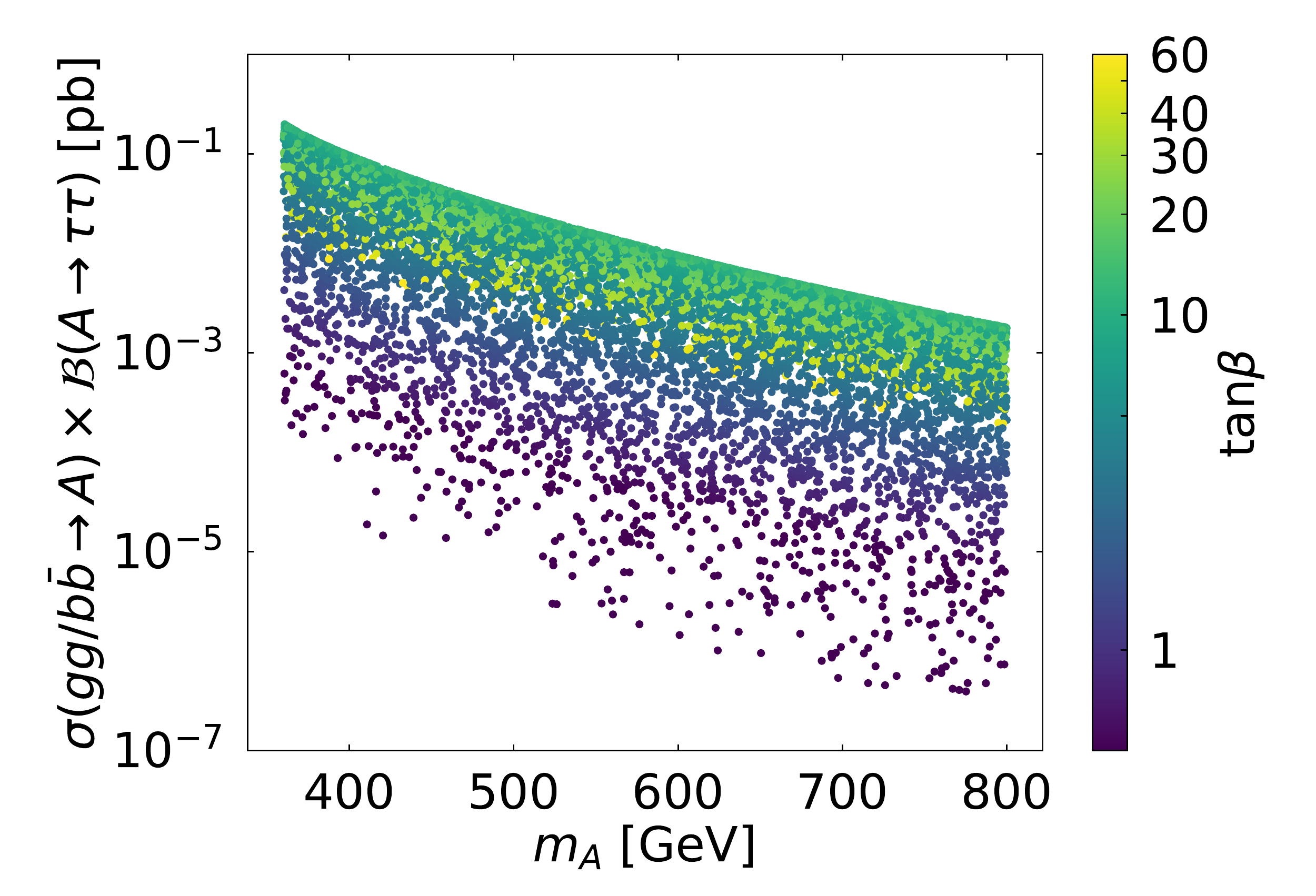}
	\label{fig:xsecscanA_typeIV}
	}
	\subfloat[][]{%
	\includegraphics[width=0.3\textwidth]{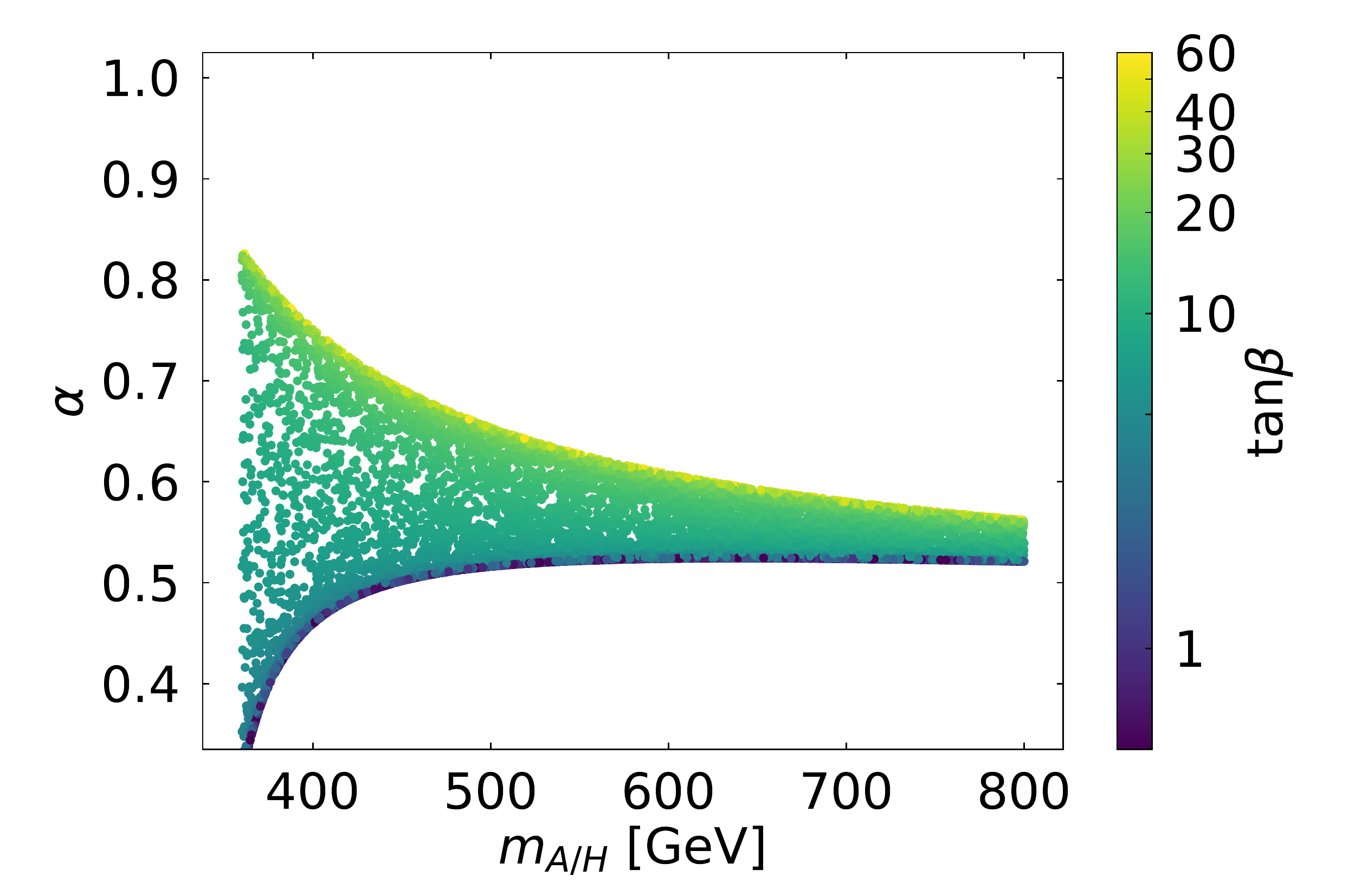}
	\label{fig:xsecscan_alpha_typeIV}
	}
	\\
	\subfloat[][]{%
	\includegraphics[width=0.3\textwidth]{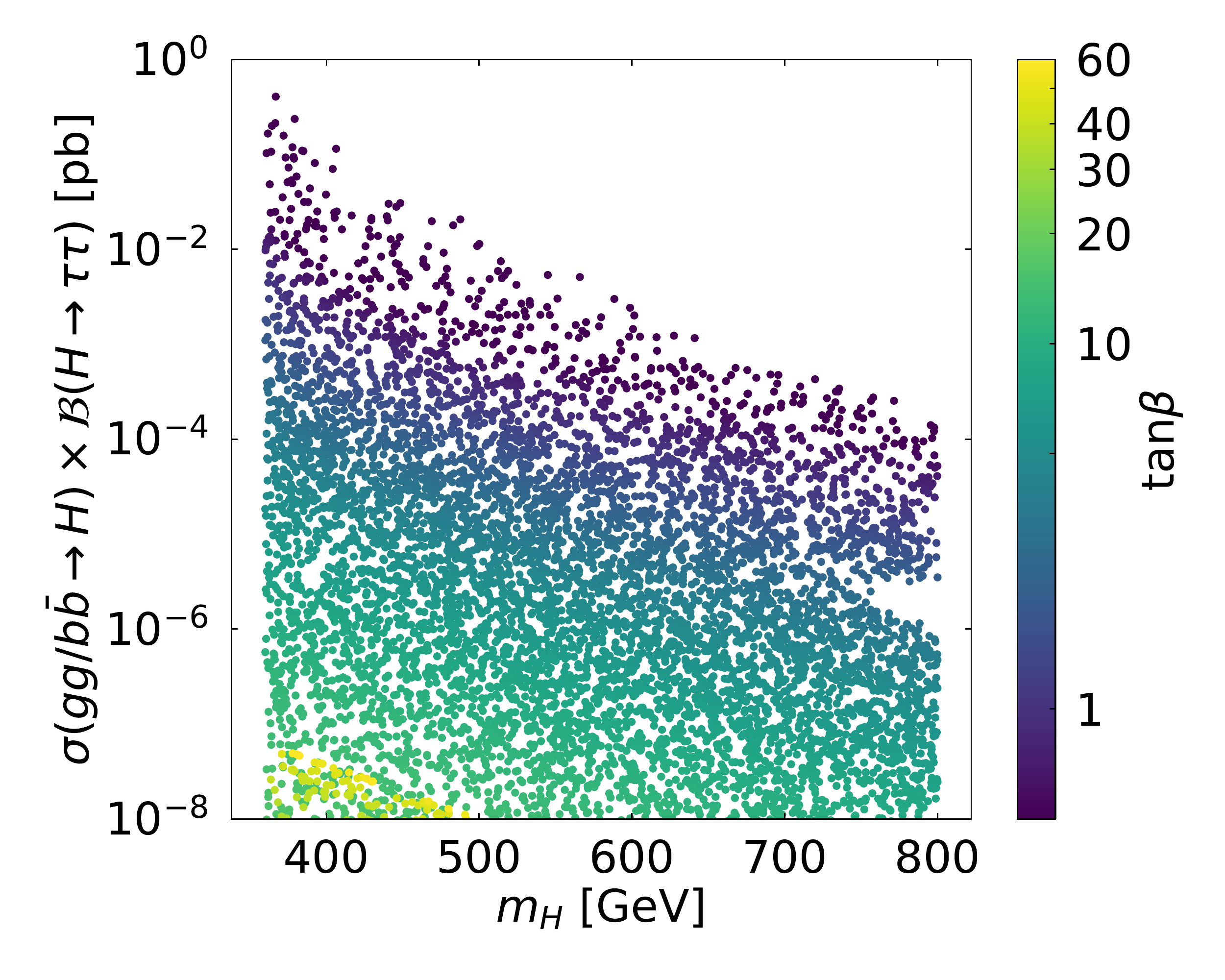}
	\label{fig:xsecscanH_typeI}
	}
	\subfloat[][]{%
	\includegraphics[width=0.3\textwidth]{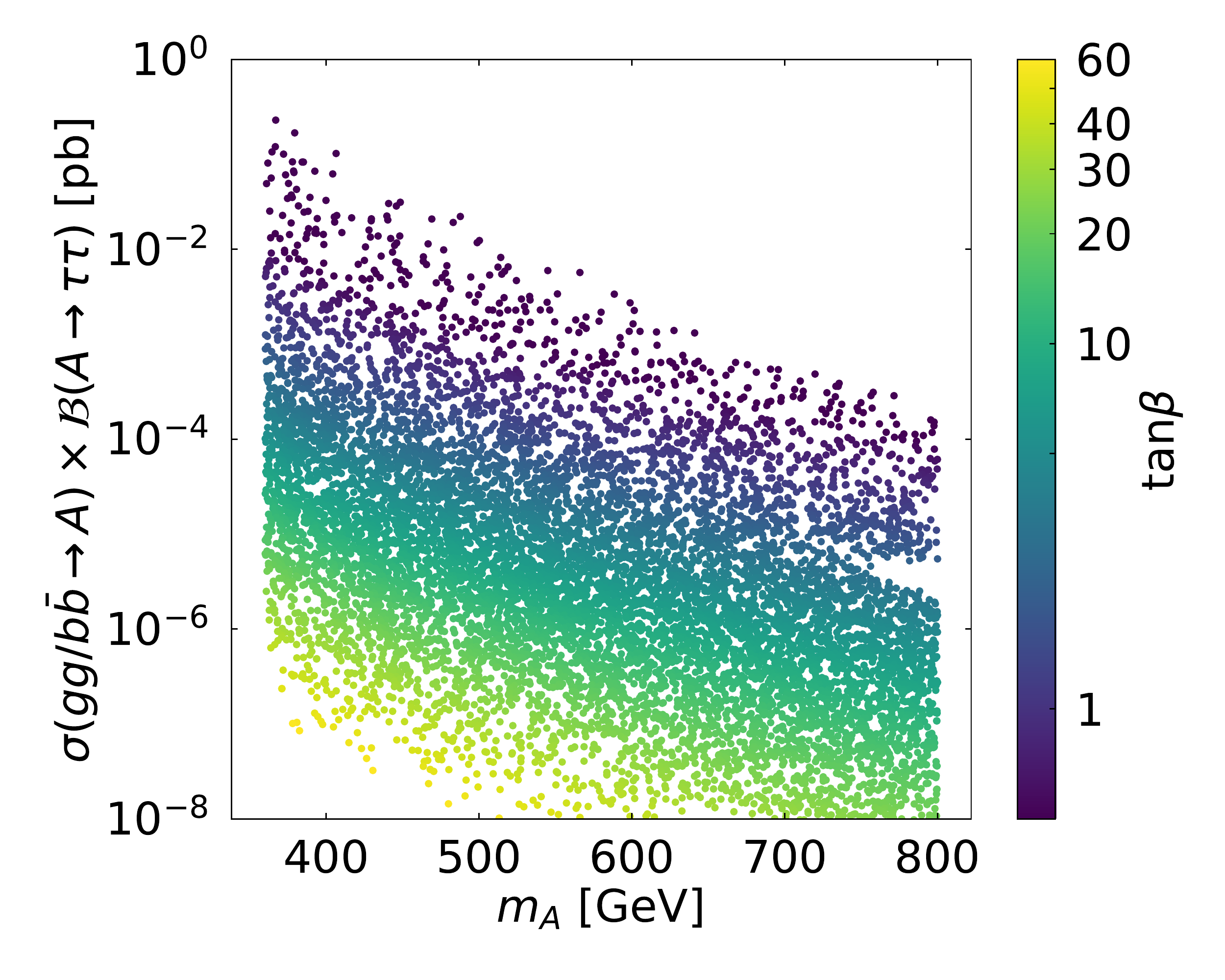}
	\label{fig:xsecscanA_typeI}
	}
	\subfloat[][]{%
	\includegraphics[width=0.3\textwidth]{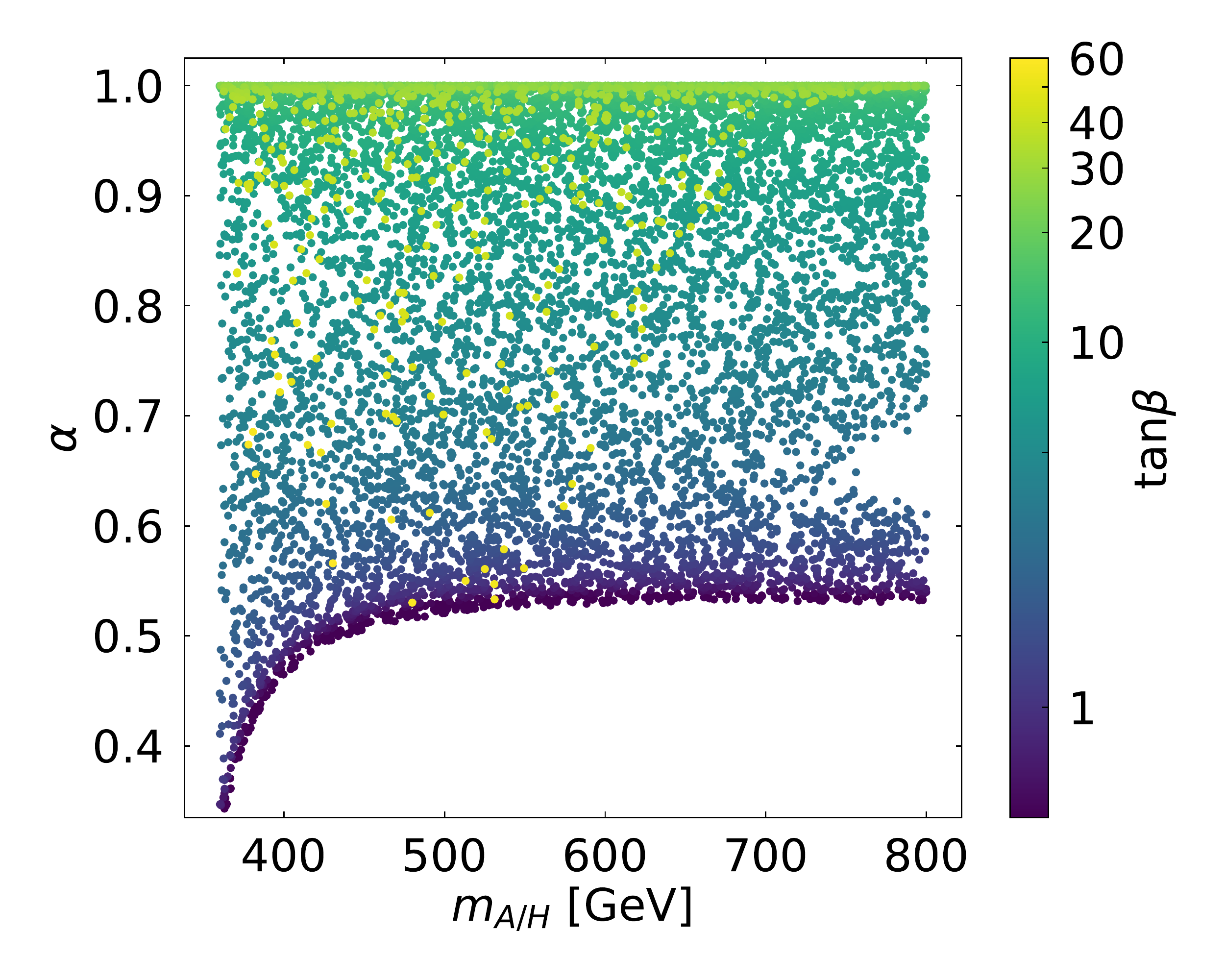}
	\label{fig:xsecscan_alpha_typeI}
	}
	\caption{\label{fig:xsecscan}\textit{Top row, lepton-specific THDM:}
	\protect\subref{fig:xsecscanH_typeIV} Signal strength for $pp
	\rightarrow H \rightarrow \tau\tau$, as a function of $m_H$ and
	$\tan\beta$. \protect\subref{fig:xsecscanA_typeIV} Similar result for
	$pp \rightarrow A \rightarrow \tau\tau$.
	\protect\subref{fig:xsecscan_alpha_typeIV} The ratio of the signal
	strength of $pp \rightarrow A \rightarrow \tau\tau$ to the total ditau
	signal strength, as defined in~\cref{eq:def_alpha}. \textit{Bottom
	row:} Corresponding results within the type-I THDM.}
\end{figure*}

%
A simple scan of the high-mass parameter regions of the (SM-aligned)
lepton-specific and type-I THDMs is performed to illustrate the parameter
dependence of the ditau signal strengths $\sigma(pp \to H) \times \mathcal{B}(H
\to \tau\tau)$ and $\sigma(pp \to A) \times \mathcal{B}(A \to \tau\tau)$, as
well as the mixture parameter $\alpha$.  The results are shown
in~\cref{fig:xsecscan}.  The parameters $m_H = m_A = m_{H^\pm}$, $\tan\beta$
and $m_{12}^2$ are varied in the scan, while we fix the light Higgs mass $m_h =
\unit[125]{GeV}$ and the neutral scalar mixing parameter
$\sin(\beta-\alpha')=1$ to ensure perfect SM alignment for the light state $h$.
The NLO cross sections are calculated with \textsf{SusHi~1.6.1}, while
branching ratios are calculated using \textsf{2HDMC~1.7.0}.  We test the
parameter points against constraints from the various collider searches for
Higgs bosons using
\textsf{HiggsBounds~4.3.1}~\cite{Bechtle:2008jh,Bechtle:2011sb,Bechtle:2013gu,Bechtle:2013wla,Bechtle:2015PMA},
while theoretical constraints are checked with \textsf{2HDMC}. Constraints from flavour
physics, in particular $\mathcal{B}(b\rightarrow s \gamma)$, disfavour
parameter regions at very low $\tan\beta$ in the type-I and lepton-specific
THDMs. These constraints were not included in the simple scan.
\end{appendices}
\begin{small}
\bibliographystyle{JHEP_pat}
\bibliography{ML_2HDM}
\end{small}
\end{document}